\documentclass[journal]{IEEEtran}
\usepackage{amsmath,amsfonts}
\usepackage{algorithmic}
\usepackage{algorithm}
\usepackage{array}
\usepackage{amssymb}
\usepackage{xhfill}
\usepackage{graphicx}
\usepackage{epstopdf}
\usepackage{float}
\usepackage{textcomp}
\usepackage{stfloats}
\usepackage{url}
\usepackage{amsthm}
\usepackage{pifont}
\usepackage{cite}
\usepackage{multirow}
\usepackage{diagbox}
\usepackage{verbatim}
\usepackage{colortbl}
\usepackage{makecell}
\def\BibTeX{{\rm B\kern-.05em{\sc i\kern-.025em b}\kern-.08em
    T\kern-.1667em\lower.7ex\hbox{E}\kern-.125emX}}
\usepackage[caption=false,font=footnotesize,labelfont=rm,textfont=rm]{subfig}
\usepackage[colorlinks,linkcolor=blue,anchorcolor=blue,citecolor=blue,urlcolor=blue]{hyperref}
\usepackage[margin=0.58in,top=0.645in,bottom=0.985in,columnsep=0.21in]{geometry}

\makeatletter
\long\def\@makecaption#1#2{%
\ifx\@captype\@IEEEtablestring%
\footnotesize\bgroup\par\centering\@IEEEtabletopskipstrut{\normalfont\small #1}\\\par\addvspace{0.25\baselineskip}{\normalfont\small\scshape #2}\par\addvspace{0.25\baselineskip}\egroup%
\@IEEEtablecaptionsepspace
\else
\@IEEEfigurecaptionsepspace
\setbox\@tempboxa\hbox{\normalfont\footnotesize {#1.}\nobreakspace\nobreakspace #2}%
\ifdim \wd\@tempboxa >\hsize%
\setbox\@tempboxa\hbox{\normalfont\footnotesize {#1.}\nobreakspace\nobreakspace}%
\parbox[t]{\hsize}{\normalfont\footnotesize\noindent\unhbox\@tempboxa#2}%
\else%
\ifCLASSOPTIONconference \hbox to\hsize{\normalfont\footnotesize\hfil\box\@tempboxa\hfil}%
\else \hbox to\hsize{\normalfont\footnotesize\box\@tempboxa\hfil}%
\fi\fi\fi}
\makeatother

\newtheorem{theorem}{Theorem}
\newtheorem{corollary}{Corollary}
\newtheorem{lemma}{Lemma}
\newtheorem{remark}{Remark}

\makeatletter
\newcommand\fs@spaceruled{\def\@fs@cfont{\bfseries}\let\@fs@capt\floatc@ruled
  \def\@fs@pre{\vspace{0.06in}\hrule height.8pt depth0pt \kern2pt}%
  \def\@fs@post{\kern2pt\hrule\kern-10pt\relax}%
  \def\@fs@mid{\kern2pt\hrule\kern2pt}%
  \let\@fs@iftopcapt\iftrue}
\makeatother

\floatstyle{spaceruled}
\restylefloat{algorithm}

\makeatletter
\renewcommand{\maketag@@@}[1]{\hbox{\m@th\normalsize\normalfont#1}}%
\makeatother

\begin{document}
\title{Multi-Static Target Position Estimation and System Optimization for Cell-Free mMIMO-OTFS ISAC}






\author{Yifei Fan,~\IEEEmembership{Graduate Student Member,~IEEE,}
		Shaochuan Wu,~\IEEEmembership{Senior Member,~IEEE,}
        \\
        Mingjun Sun, Lin Huo, Jianchao Su, and 
        Haojie Wang,~\IEEEmembership{Student Member,~IEEE}

        \vspace{-2pt}

\thanks{This work was supported by the National Natural Science Foundation of China under Grant 62271167. (\emph{Corresponding author: Shaochuan Wu.})}

\thanks{ {The authors are with the School of Electronics and Information Engineering, Harbin Institute of Technology, Harbin 150001, China (e-mail: yifan@\parfillskip=0pt\par} 
\noindent stu.hit.edu.cn; scwu@hit.edu.cn; sunmj@stu.hit.edu.cn; lhuo@stu.hit.edu.cn; sujianchao@stu.hit.edu.cn; whj2@stu.hit.edu.cn). Parts of this paper have been accepted by the 2025 IEEE GLOBECOM conference~\cite{Fan2025Multi}.}

}

\maketitle

\begin{abstract}
This paper investigates multi-static position estimation in cell-free massive multiple-input multiple-output (CF mMIMO) architectures, where orthogonal time frequency space (OTFS) is used as an integrated sensing and communication (ISAC) signal. A maximum likelihood position estimation scheme is proposed, where the required search space is reduced by employing a common reference system. Closed-form expressions for the Cram\'{e}r-Rao lower bound and the position error bound (PEB) in multi-static position estimation are derived, providing quantitative evaluations of sensing performance. These theoretical bounds are further generalized into a universal structure to support other ISAC signals. To enhance overall system performance and adapt to dynamic network requirements, a joint AP operation mode selection and power allocation algorithm is developed to maximize the minimum user communication spectral efficiency (SE) while ensuring a specified sensing PEB requirement. Moreover, a decomposition method is introduced to achieve a better tradeoff between complexity and ISAC performance. The results verify the effectiveness of the proposed algorithms, demonstrating the superiority of the OTFS signal through a nearly twofold SE gain over the orthogonal frequency division multiplexing (OFDM) signal. These findings highlight promising advantages of the CF-ISAC systems from a novel parameter estimation perspective, particularly in high-mobility vehicle-to-everything applications.

\end{abstract}


\section{Introduction}
\IEEEPARstart{I}{ntegrated} sensing and communication (ISAC) has emerged as a key enabling technology in the forthcoming 6G era, as envisioned in the IMT-2030 framework~\cite{Luo2025ISAC}. By leveraging shared resources and co-designing communication and sensing (C\&S) functionalities, ISAC systems can achieve coordination gains and enhance overall performance while reducing hardware costs and alleviating spectrum congestion, representing a more profound integration paradigm~\cite{Lu2024Integrated,liu2022integrated}. Under this vision, the existing cellular networks are expected to be equipped with ubiquitous perceptive capability, evolving into perceptive mobile networks~\cite{zhang2021enabling}. Based on the spatial distribution of the transmitter and sensing receiver, these network-attached ISAC systems are categorized into mono-static, bi-static, and multi-static sensing configurations~\cite{Liu2023Seventy}.

In an ISAC system based on cellular networks, the transmitter and receiver are typically co-located and both functions are performed by cellular access points (APs), characterizing a mono-static sensing configuration~\cite{GonzalezPrelcic2024integrated}. However, the single observation angle generated by the cellular AP can be easily blocked in complex propagation environments~\cite{Liu2024Cooperative}. In addition, the cellular networks often suffer from a fairness problem at cell edges, resulting in unreliable C\&S services for ultra-reliable applications, such as vehicle-to-everything (V2X)~\cite{Mohammadi2024Next}. To address the limitations inherent in cellular networks, cell-free massive multiple-input multiple-output (CF mMIMO) has emerged as a promising solution~\cite{Zhang2020Prospective,chen2022survey}. As a representative of multi-static sensing, the CF mMIMO architecture enables multi-angle observations and achieves spatial diversity by leveraging geographically distributed transmitters and receivers~\cite{Mao2024Communication,Mohammadi2023Network}. Moreover, the cell edges can be completely eliminated in this architecture by ensuring uniform and ubiquitous service~\cite{ammar2021user}.

\subsection{Related Work}
Driven by their significant potential, attaching the ISAC capabilities into the CF mMIMO network architecture has attracted growing research interest~\cite{behdad2024multi,Elfiatoure2025Multiple,Sakhnini2025Distributed,Zeng2025Multi,Pucci2025Cooperative,Zhang2025Target,Tang2025Cooperative}. Specifically, a power allocation algorithm was proposed in~\cite{behdad2024multi} to maximize detection probability while meeting communication signal-to-interference-plus-noise ratio (SINR) constraints. A similar problem was studied in~\cite{Elfiatoure2025Multiple}, where mainlobe-to-average-sidelobe ratio (MASR) is used as the sensing metric and a penalty-based method was developed to solve a joint operation mode selection and power control design problem. The authors of~\cite{Sakhnini2025Distributed} studied a distributed radar and communication system operating in the uplink, where interference cancellation and power control are employed to mitigate the impact of user interference on radar. This work was further extended to network-assisted full-duplex CF networks to support asymmetric uplink/downlink communication requirements~\cite{Zeng2025Multi}. Additionally, in~\cite{Pucci2025Cooperative}, target position estimation was investigated in cooperative ISAC systems, where a general maximum likelihood (ML) framework was derived. A two-stage target localization scheme based on 5G NR orthogonal frequency division multiplexing (OFDM) signals was proposed in~\cite{Zhang2025Target}, where ill-conditioned measurements are effectively eliminated. A two-stage scheme was proposed in~\cite{Tang2025Cooperative} for cooperative ISAC systems, where a spatial smoothing tensor decomposition scheme was introduced to estimate the targets’ parameters, and a false removing minimum spanning tree (MST)-based data association method was developed to fusion the positions and true velocities of the targets.

The above works have been based on the application of traditional integrated signals. However, given the sensitivity of the OFDM signal to Doppler shifts in high-mobility applications such as V2X, the emerging orthogonal time frequency space (OTFS) signal stands out as a superior candidate for the CF-ISAC systems~\cite{li2021performance}. By modulating information symbols in the delay-Doppler (DD) domain, the OTFS signal exhibits robustness against delay and Doppler spreads~\cite{Gong2023Simultaneous}. Moreover, the OTFS signal can effectively capture the range and velocity characteristics of moving targets, achieving parameter estimation accuracy comparable to specialized radar signals, outperforming the OFDM signal~\cite{gaudio2020effectiveness}.

Recently, a few studies have taken a step further towards employing the OTFS signal in the CF-ISAC systems~\cite{Fan2024Power,Singh2025Target,Das2025Low}. Particularly, the authors of~\cite{Fan2024Power} derived a closed-form spectral efficiency (SE) expression regarding optional sensing beams. A power allocation strategy was proposed to maximize the minimum communication SINR between users while guaranteeing a specified sensing SINR value. Further, in~\cite{Singh2025Target}, target detection performance was evaluated in a sensing-centric approach, where transmit power was optimized to maximize the sensing signal-to-noise ratio (SNR) while ensuring a required communication quality-of-service (QoS). The authors of~\cite{Das2025Low} incorporated a low-complexity precoding scheme for the CF-ISAC system, using the MASR metric to evaluate sensing performance.

\begin{table*}[!ht]
\small
    \centering
    \captionsetup{font=normalsize}
    \caption{Contrasting the Contributions of This Paper to the CF-ISAC Literature}
    \label{tab1}
    \renewcommand\arraystretch{1.2}{
    \begin{tabular}{|c|c|c|c|c|c|c|c|c|}
    \hline
        \textbf{Contributions} & \!\textbf{This paper}\! &\!\!\!\cite{behdad2024multi}-2024\!\! &\!\!\!\cite{Elfiatoure2025Multiple}-2025\!\! &\!\!\!\cite{Sakhnini2025Distributed}-2025\!\! &\!\!\!\cite{Zeng2025Multi}-2025\!\! &\!\!\!\cite{Fan2024Power}-2024\!\! &\!\!\!\cite{Singh2025Target}-2025\!\! &\!\!\!\cite{Das2025Low}-2025\!\! \\ \hline
        OTFS signal                 & \checkmark & ~          & ~          & ~          & ~          & \checkmark & \checkmark & \checkmark  \\ \hline
        AP mode selection           & \checkmark & ~          & \checkmark & \checkmark & \checkmark & ~          & ~          & ~           \\ \hline
        \textbf{AP array impact}    & \checkmark & ~          & ~          & ~          & ~          & ~          & ~          & ~           \\ \hline
        \textbf{Position estimation}& \checkmark & ~          & ~          & ~          & ~          & ~          & ~          & ~           \\ \hline
        \textbf{CRLB analysis}      & \checkmark & ~          & ~          & ~          & \checkmark & ~          & ~          & ~           \\ \hline
        Performance optimization    & \checkmark & \checkmark & \checkmark & \checkmark & \checkmark & \checkmark & \checkmark & ~           \\ \hline
    \end{tabular}}
    \vspace{-3pt}
\end{table*}

\subsection{Contributions}
Sensing tasks mainly involve \emph{target detection} and \emph{parameter estimation}~\cite{Lu2024Integrated}. The aforementioned studies have focused on \emph{target detection} by constraining or maximizing the sensing SINR, while leaving the latter task — \emph{parameter estimation} unexplored. To fill this gap, this paper investigates multi-static position estimation, using the position error bound (PEB) as a sensing performance metric. To the best of the authors’ knowledge, the \emph{parameter estimation} performance of targets in the CF
mMIMO-OTFS ISAC systems has not been explored in the existing literature. The main contributions of this paper are contrasted in Table~\ref{tab1} and further summarized as follows.
\begin{itemize}
  \item[$\bullet$] Considering the antenna array directions of the APs, a comprehensive CF-ISAC system model is developed based on the OTFS signal. Building on this model, we propose a general ML multi-static position estimation scheme, along with a simplified version for widely-separated targets. By defining a common reference system, the proposed scheme avoids complex coordinate transformations, enabling efficient operation within a smaller search space, thereby significantly reducing computational complexity.
  \item[$\bullet$] To evaluate the proposed ML position estimation scheme, closed-form expressions for the Cramér-Rao lower bound (CRLB) and PEB are derived based on the OTFS signal, serving as sensing performance metrics. We further establish a universal CRLB structure for multi-static sensing and demonstrate the compatibility of these theoretical bounds with other ISAC signals, using the OFDM signal as an example. To facilitate efficient analysis and optimization of the position estimation performance, this study introduces a low-complexity PEB approximation and provides detailed conditions to ensure its accuracy. 
  \item[$\bullet$] A joint problem of AP operation mode selection and power allocation is formulated, considering per-AP power constraints and a PEB constraint for multi-static position estimation. To solve this challenging mixed-integer non-convex problem, we reformulate it into a more tractable continuous-variable problem, which is then addressed via successive convex approximation (SCA) techniques. Alternatively, a low-complexity decomposition method is developed, comprising a distance-based AP mode selection algorithm and subsequent power allocation optimization for the selected AP mode.
  \item[$\bullet$] The results validate the effectiveness of the proposed ML position estimation scheme and the accuracy of the derived PEB expressions, intuitively illustrating the coordination gain of ISAC in sensing. In addition, we provide an in-depth analysis of the significant impact of the AP antenna array directions on sensing performance.
\end{itemize}

$\emph{Notation:}$ Lowercase letters, boldface lowercase letters, and boldface uppercase letters denote scalars, column vectors, and matrices, respectively. The superscripts $\left( \cdot \right)^*$, $\left( \cdot \right)^{\mathrm{T}}$, $\left( \cdot \right)^{-1}$, and $\left( \cdot \right)^{\dagger}$ represent the conjugate, transpose, inverse, and conjugate-transpose operations, respectively. The operators $\mathrm{Tr\left( \cdot \right)}$, $\mathbb{E}\left\{ \cdot \right\}$, $\odot$, and $\otimes$ denote the trace, expectation, Hadamard product, and Kronecker product, respectively; $\left\lceil\cdot\right\rceil$ is the ceiling function, and $\operatorname{diag}\{\cdot\}$ returns a diagonal matrix. Finally, $\left\| \cdot  \right\|$ and $\left| \cdot \right|$ represent the vector and scalar Euclidean norms, respectively.

\section{System Model}
\begin{figure}[!t]
  \centering
  \includegraphics[width=2.5in]{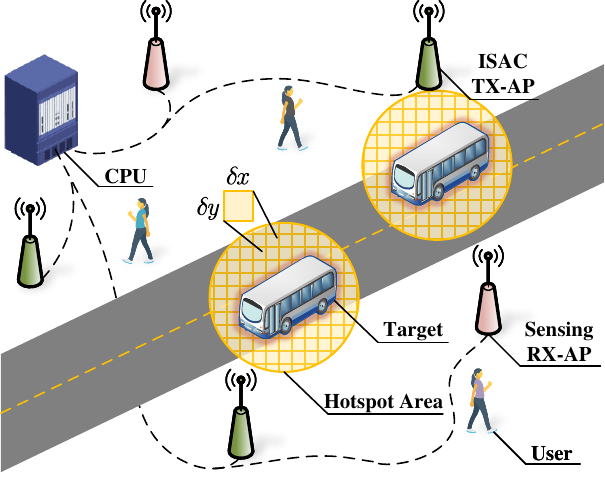}
  \vspace{-3mm}
  \caption{Illustration of the multi-static CF-ISAC system setup.}
  \label{fig:fig_1}
  \vspace{-12.5pt}
\end{figure}

This study considers a multi-static CF-ISAC system during the downlink phase, where the OTFS is used as an integrated signal. All $N_{\mathrm{AP}}$ APs are connected to a centralized processing unit (CPU) synchronously, and each AP is equipped with a uniform linear array (ULA) of $M_{\mathrm t}$ antennas. As depicted in Fig.~\ref{fig:fig_1}, each AP functions either as an ISAC transmitter or a sensing receiver, determined by a designed mode selection scheme. The $N_{\mathrm{tx}}$ transmitting APs employ maximum-ratio (MR) precoding to transmit integrated signals, jointly serving $K_{\mathrm u}$ single-antenna users while sensing $T_{\mathrm{g}}$ targets. The remaining $N_{\mathrm{rx}}$ receiving APs then collect echo signals to estimate the targets' positions.\footnote{To clearly present the multi-static sensing model, the indices of the transmitting and receiving APs are denoted by $p=1,\ldots,N_{\mathrm{tx}}$ and $r=1,\ldots,N_{\mathrm{rx}}$, respectively. The AP mode selection schemes will be introduced in Section~\ref{sec:Problem Formulation and Proposed Solutions}.} Assuming a hotspot area has been identified during a prior target detection phase, the search space for target positions is consequently configured within the hotspot area~\cite{Pucci2025Cooperative}.


\begin{table*}[!ht]
\begin{center}
\caption{List of Notations}
\label{tab0}
\renewcommand\arraystretch{1.5}{
\begin{tabular}{ c  c || c  c }
\Xhline{0.8pt}
\textbf{Symbol} & \textbf{Definition} & \textbf{Symbol} & \textbf{Definition}\\
\Xhline{0.6pt}
$N_{\mathrm{AP}}$ & The total number of APs & 
$N_{\mathrm{tx}}$, $N_{\mathrm{rx}}$ & Number of transmitting and receiving APs, respectively\\
\hline
$K_{\mathrm u}$ & Number of users &  
$T_{\mathrm g}$ & Number of targets \\
\hline 
$M$ & Number of subcarriers &  
$N$ & Number of symbols \\
\hline 
$\Delta f$ & Subcarrier bandwidth &  
$T$ & Symbol duration \\
\hline
$M_{\mathrm t}$ & Number of antennas per AP & 
$P_{p}$ & Downlink transmit power at transmitting AP $p$ \\
\hline 
$\eta_{pq}, \eta_{pv}$ & \makecell{Power allocation coefficient at the $p$th transmitting AP \\ for the $q$th user and the $v$th target, respectively}  & 
$\tau _{pq,i}, \nu _{pq,i}\!$ & \makecell{Delay and Doppler shift of the $i$th path from the $p$th \\ transmitting AP to the $q$th user, respectively} \\
\hline 
$\tau _{p,r,v}, \nu _{p,r,v}\!$ & \makecell{Bi-static delay and Doppler shift from transmitting AP \\ $p$ via the $v$th target to receiving AP $r$, respectively}  & 
$\beta_{p,r,v}$ & \makecell{Sensing channel gain from the $p$th transmitting AP \\ via the $v$th target to the $r$th receiving AP} \\
\hline 
$\omega_{p,r,v}^{\mathrm{r}}$ & AoA between the $v$th target and the $r$th receiving AP  & 
$\omega_{p,r,v}^{\mathrm{t}}$ & AoD between the $v$th target and the $p$th transmitting AP  \\
\hline
$\gamma_{\mathrm s}$ & The maximum sensing CRLB threshold & 
$L_{pq}$ & Number of paths from transmitting AP $p$ to user $q$ \\
\hline 
$\mathbf{x}_{q}, \mathbf{x}_{v}$ & \makecell{Transmitted DD domain signal for the $q$th \\ user and the $v$th target, respectively} & 
$\mathbf{y}_{q}, \mathbf{y}_{r}$ & \makecell{Received DD domain signal at the $q$th user \\ and the $r$th receiving AP, respectively}\\
\hline 
$\mathbf{p}_v, \mathbf{p}_p, \mathbf{p}_r$ & \makecell{The positions of the $v$th target, the $p$th transmitting \\ AP and the $r$th receiving AP, respectively} &
$\mathbf{u}_p, \mathbf{u}_r$ & \makecell{The unit direction vectors of the antenna elements at the \\ $p$th transmitting AP and the $r$th receiving AP, respectively} \\
\hline 
$\mathbf{h}_{pq,i}, \hat{\mathbf{h}}_{pq,i}$ & \makecell{Channel impulse response and its estimate of the $i$th path \\ from the $p$th transmitting AP to the $q$th user, respectively} & 
$\mathbf{H}_{pq}, \hat{\mathbf{H}}_{pq}$ & \makecell{Effective DD domain channel and its estimate from \\ the $p$th transmitting AP to the $q$th user, respectively} \\
\hline 
$\hat{\mathbf{h}}_{pv}$ & \makecell{The sensing precoding vector at the $p$th \\ transmitting AP for the $v$th target} & 
$\hat{\mathbf{H}}_{pv}$ & \makecell{The DD domain sensing precoding matrix at \\ the $p$th transmitting AP for the $v$th target} \\
\hline 
${\boldsymbol{\Psi}} _{pq}^{(i)}, \boldsymbol{\Psi}_{p,r,v}\!$ & \makecell{DD domain index matrix of the $i$th communication path \\ and sensing reflected path via the $v$th target, respectively}& 
$\mathbf{H}_{prv}$ & \makecell{Sensing reflected channel from the $p$th transmitting AP \\ via the $v$th target to the $r$th receiving AP} \\
\Xhline{0.8pt}
\end{tabular}}
\end{center}
\vspace{-9pt}
\end{table*}

\subsection{Downlink Communication Model}
The OTFS signal is assumed to have $M$ subcarriers with a subcarrier spacing of $\Delta f$, and $N$ symbols with a symbol duration of $T$. A cyclic prefix (CP) of sample length $N_{\mathrm{cp}}$ is added to each block, ensuring the corresponding CP duration is larger than the maximum delay spread, i.e., $T_{\mathrm{cp}}\geq\tau_{\max}$. The information symbols for the $q$th user ${x_q[k,\ell]}$ are scheduled on the DD grid $\Gamma=\big\{\frac{k}{NT},\frac{\ell}{M\Delta f}\big\}$, where $k$ and $\ell$ represent the Doppler and delay indexes, respectively. After performing the inverse symplectic finite Fourier transform (ISFFT), the DD domain symbols $x_q[k,\ell]$ are converted to time-frequency (TF) domain as follows:
\vspace{-2mm}
\begingroup\makeatletter\def\f@size{9.5}\check@mathfonts
\begin{equation} X_{q}[n,m] = \frac {1}{\sqrt {MN}}\sum _{k=0}^{N-1}\sum _{\ell =0}^{M-1} x_{q}[k,\ell] e^{j2\pi \left ({\frac {nk}{N}-\frac {m\ell }{M}}\right)},\end{equation} 
\endgroup
\vspace{-1mm}
\vspace{-1mm}

\noindent for $n,k=0,\ldots,N-1$ and $m,l=0,\ldots,M-1$. Further, by performing the Heisenberg transform, $X_{q}[n,m]$ are converted to a time-domain signal as
\vspace{-1.25mm}
\begingroup\makeatletter\def\f@size{9.5}\check@mathfonts
\begin{align} 
\hspace {-0.4pc}s_{pq}(t)=\sum_{n=0}^{N-1}{\sum_{m=0}^{M-1}}\sqrt{\eta _{pq}}{X_q}[n,m]g_{tx}(t-nT)e^{j2\pi m\Delta f(t-nT)},
\label{eq:s_t}
\end{align}
\vspace{-1.25mm}
\vspace{-1mm}

\noindent where $g_{tx}(t)$ is the transmitting pulse-shaping filter, $\eta _{pq},\, p=1,\ldots,N_{\mathrm{tx}},\,q=1,\ldots,K_{\mathrm{u}}$ are the power allocation coefficients set to make the average transmit power $P_p$ at each transmitting AP satisfy the following power constraint 
\vspace{-1mm}
\begingroup\makeatletter\def\f@size{9}\check@mathfonts
\begin{equation} 
P_p=\mathbb {E}\left \{{\Big| \sum\nolimits _{q=1}^{K_{\mathrm{u}}}s_{pq}(t)+\sum\nolimits _{v=1}^{T_{\mathrm{g}}}s_{pv}(t)\Big|^{2}}\right \} \leq P_{\mathrm{d}},\label{eq:power_constraint}\end{equation}
\endgroup

\noindent where $s_{pv}(t)$ is the time-domain sensing signal for target $v$ obtained by applying a similar procedure as in~\eqref{eq:s_t}, and $P_{\mathrm{d}}$ denotes the maximum downlink transmit power.

Considering the doubly selective fading caused by high user mobility, the DD domain channel impulse response from transmitting AP $p$ to user $q$ can be expressed as~\cite{Fan2024Power,Zegrar2024OTFS}
\vspace{-0.5mm}
\begingroup\makeatletter\def\f@size{9.5}\check@mathfonts
\begin{equation} \mathbf{h}_{pq}(\tau,\nu) = \sum\nolimits _{i=1}^{ L_{pq}} \mathbf{h}_{pq,i}\delta (\tau - \tau _{pq,i})\delta (\nu - \nu _{pq,i}),\label{eq:h_pq}\end{equation}
\endgroup
\vspace{-1mm}
\vspace{-1mm}
\vspace{-1mm}
\vspace{-0.5mm}

\noindent where the channel vector $\mathbf{h}_{pq,i}\sim \mathcal{CN}(\mathbf{0},\mathbf{R}_{pq,i})$ follows a correlated Rayleigh fading model, with its spatial correlation matrix $\mathbf{R}_{pq,i}=\mathbb{E}\{\mathbf{h}_{pq,i}^{\phantom{\dagger}}\mathbf{h}_{pq,i}^{\dagger}\}\in \mathbb{C}^{M_{\mathrm t}\times M_{\mathrm t}}\vphantom{\underline{\overline{\hat{\mathbf{R}}^{\dagger}_{pq,i}}}}$ reflecting the combined effect of geometric path loss, shadowing, and spatial correlation between antennas~\cite{demir2021foundations}. Moreover, $\tau_{pq,i}$, $\nu_{pq,i}$, and $L_{pq}$ represent the $i$th path's delay, Doppler shift, and the number of paths from transmitting AP $p$ to user $q$, respectively.




For downlink communication, the transmitting APs apply MR precoding to transmit integrated signals to serve $K_{\mathrm{u}}$ users~\cite{AbuShaban2018Error,Zeng2025Multi,Ren2025Two}. The signal received at the $q$th user is given by
\vspace{-0.5mm}
\begingroup\makeatletter\def\f@size{9.3}\check@mathfonts
\begin{align}
\hspace {-0.4pc}r_{q}(t) =&\nonumber \int_\tau \int_\nu \sum\nolimits_{p=1}^{N_{\mathrm{tx}}} \mathbf{h}_{pq}^\mathrm{T}(\tau,\nu)\bigg( \sum\nolimits _{q^\prime=1}^{K_{\mathrm{u}}}\hat{\mathbf{h}}_{pq^\prime}^{*}(\tau,\nu)s_{pq^\prime}(t-\tau)  \\
&\!\!+\!\sum\nolimits _{v=1}^{T_{\mathrm{g}}}\!\hat{\mathbf{h}}_{pv}^{*}(\tau,\nu)s_{pv}(t-\tau) \!\bigg)e^{j2\pi \nu (t-\tau)} d\tau d\nu + w_{q}(t),\!\!
\end{align}
\endgroup
\vspace{-1mm}
\vspace{-1mm}
\vspace{-1mm}
\vspace{-0.5mm}

\noindent where $\hat{\mathbf{h}}_{pv}$ is the sensing precoding vector given by~\eqref{eq:sensing_BF}, $\hat{\mathbf{h}}_{pq}$ is a unit-norm estimate of channel vector $\mathbf{h}_{pq}$, and $w_{q}(t)\sim\mathcal{CN}(0,\sigma_w^2)$ is the noise received by the $q$th user. After performing the Wigner transform equipped with a receiving filter $g_{rx}(t)$, the TF domain received samples are obtained by a sampler as
\vspace{-0.5mm}
\begingroup\makeatletter\def\f@size{9.5}\check@mathfonts
\begin{equation} 
Y_{q}[n,m]=\int r_{q}(t) g_{rx}(t-nT) e^{-j2\pi m \Delta f (t-nT)} dt.
\end{equation}
\endgroup
\vspace{-1mm}
\vspace{-1mm}
\vspace{-1mm}

{\noindent Finally, by applying the SFFT to $Y_{q}[n,m]$ and assuming non-ideal rectangular windows are used in the transmitting and receiving pulse-shaping filters, the DD domain signal received at the $q$th user can be formulated in a vector form as
\vspace{-1mm}
\begingroup\makeatletter\def\f@size{9.3}\check@mathfonts
\begin{align} 
\mathbf{y}_{q}=&\underbrace{\sum _{p=1}^{N_{\mathrm{tx}}} \eta _{pq}^{\frac{1}{2}} {\mathbf{H}}_{pq} \hat {\mathbf{H}}_{pq}^{\dagger} {\mathbf{x}} _{q}}_{\text{Desired signal}} \nonumber + \underbrace{\sum _{p=1}^{N_{\mathrm{tx}}} \sum _{q'\neq q}^{K_{\mathrm{u}}} \eta _{pq'}^{\frac{1}{2}}{\mathbf{H}}_{pq}\hat {\mathbf{H}}_{pq'}^{\dagger} {\mathbf{x}} _{q'}}_{\text{Inter-user interference}} \\[-4pt]
&+\, \underbrace{\sum _{p=1}^{N_{\mathrm{tx}}} \sum _{v=1}^{T_{\mathrm{g}}} \eta _{pv}^{\frac{1}{2}}{\mathbf{H}}_{pq}\hat {\mathbf{H}}_{pv}^{\dagger} {\mathbf{x}} _{v}}_{\text{Sensing interference}} + \underbrace{\vphantom{\sum _{p=1}^{N_{\mathrm{tx}}}}{\mathbf{w}}_{q}}_{\text{Noise}},\label{eq:y_ps}\end{align}
\endgroup
\vspace{-1mm}
\vspace{-1mm}


\begingroup\makeatletter\def\f@size{9.3}\check@mathfonts
\noindent where ${\mathbf{w}}_{q} \sim \mathcal{C N}\left(\mathbf{0}_{MN}, \sigma_w^2 \mathbf{I}_{MN}\right)$ is the noise vector at user~$q$, and the effective DD domain channel between the $p$th transmitting AP and the $q$th user is given by~\cite{Dehkordi2023Beam}
\vspace{-0.5mm}
\begin{equation} {\mathbf{H}}_{pq} = \sum\nolimits _{i=1}^{ L_{pq}} \left( \mathbf h_{pq,i}^\mathrm{T}\otimes {\boldsymbol{\Psi}} _{pq}^{(i)} \right)\in \mathbb{C}^{MN\times M_{\mathrm t}MN},\label{eq:DD_channel}\end{equation}
\endgroup

\noindent where $\boldsymbol{\Psi}\in \mathbb{C}^{MN\times MN}$ contains channel delay and Doppler information. By defining $l_{\tau}\triangleq\lceil\tau M\Delta f\rceil$, the elements of $\boldsymbol{\Psi}$ are obtained by~\eqref{eq:Psi}, which is shown at the bottom of the page.
\stepcounter{equation}

\addtocounter{equation}{-1}
\begin{figure*}[hb] 
\vspace{-3mm}
\xhrulefill {black}{1pt}
  \centering
    \begin{small}
    \begin{equation}
        \Psi_{k,k^{\prime},l,l^{\prime}}\approx\frac{1}{NM}\!\sum_{n^{\prime}=0}^{N-1}\underbrace{\vphantom{\begin{array}{ll}1&l\in\mathcal{L}_{\mathrm{ICI}}(\tau):=[0,M-l_{\tau}-1]\\\\e^{-j2\pi\left(\nu T+\frac{k}{N}\right)}&l\in\mathcal{L}_{\mathrm{ISI}}(\tau):=[M-l_{\tau},M-1]\end{array}}e^{j2\pi(k^{\prime}-k+\nu NT)\frac{n^{\prime}}{N}}}_{\triangleq\alpha_{n^{\prime},k,k^{\prime}}(\nu)}
        \sum_{m^{\prime}=0}^{M-1}\underbrace{e^{j2\pi(l^{\prime}-l+\tau M\Delta f)\frac{m^{\prime}}{M}}e^{j2\pi\nu\frac{l^{\prime}}{M\Delta f}}\left\{\begin{array}{ll}1&l^{\prime}\in\mathcal{L}_{\mathrm{ICI}}(\tau):=[0,M-l_{\tau}-1]\\\\e^{-j2\pi\left(\nu T+\frac{k^{\prime}}{N}\right)}&l^{\prime}\in\mathcal{L}_{\mathrm{ISI}}(\tau):=[M-l_{\tau},M-1]\end{array}\right.\!\!\!}_{\triangleq\beta_{m^{\prime},k^{\prime},l,l^{\prime}}(\nu,\tau)}
        \label{eq:Psi}
    \end{equation}
    \end{small}
\end{figure*}
\addtocounter{equation}{0}

\subsection{Multi-Static Sensing Model}
{Suppose target $v$ is located at $\mathbf{p}_v=[x_v,y_v]^{\mathrm{T}}$ in the horizontal coordinate with a velocity of $\mathbf{v}_v$. Similarly, let $\mathbf{p}_p$ and $\mathbf{p}_r$ denote the positions of the $p$th transmitting AP and the $r$th receiving AP, respectively. Then, the parameters of the reflected path from transmitting AP $p$ via the $v$th target to receiving AP $r$, namely the angle of arrival (AoA), angle of departure (AoD), bi-static \parfillskip=0pt\par}

\begin{figure}[H]
  \centering
  \includegraphics[width=2.3in]{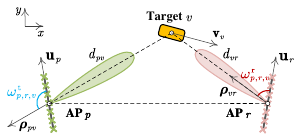}
  \vspace{-3mm}
  \caption{Illustration of the bi-static reflected path parameters.}
  \label{fig:fig_12}
  \vspace{-9pt}
\end{figure}

\noindent delay and Doppler, $\boldsymbol{\theta}_{p,r,v}^{(1)}\!\triangleq[\omega_{p,r,v}^{\mathrm{r}},\omega_{p,r,v}^{\mathrm{t}},\tau_{p,r,v},\nu_{p,r,v}]^\mathrm{T}\!,$ can be obtained by
\vspace{-1mm}
\vspace{-1mm}
\vspace{-0.5mm}
\begingroup\makeatletter\def\f@size{9.5}\check@mathfonts
\begin{equation}
\begin{aligned}
\hspace {1.5pc}&\omega_{p,r,v}^{\mathrm{r}}=\pi\boldsymbol{\rho}_{vr}^{\mathrm{T}}\mathbf{u}_r^{\phantom{0}},\quad\tau_{p,r,v}=\left(d_{pv}+d_{vr}\right)/c,\\
&\omega_{p,r,v}^{\mathrm{t}}=\pi\boldsymbol{\rho}_{pv}^{\mathrm{T}}\mathbf{u}_p^{\phantom{0}},\quad\nu_{p,r,v}=\mathbf{v}_v^{\mathrm{T}}\left(\boldsymbol{\rho}_{pv}+\boldsymbol{\rho}_{vr}\right)/\lambda_c,
\label{eq:path_parameters}
\end{aligned}
\end{equation}
\endgroup
\vspace{-1mm}
\vspace{-1.5mm}

\noindent where $\lambda_{c}$ is the carrier wavelength; $\mathbf{u}_p$ and $\mathbf{u}_r$ represent the unit direction vectors of the antenna elements at transmitting AP $p$ and receiving AP $r$, respectively~\cite{Gong2023Simultaneous}; $d_{pv}=\|\mathbf{p}_p-\mathbf{p}_v\|$ and $d_{vr}=\|\mathbf{p}_v-\mathbf{p}_r\|$ denote the distances from transmitting AP $p$ and receiving AP $r$ to the $v$th target location, respectively; and unit vectors $\boldsymbol{\rho}_{pv}=(\mathbf{p}_p-\mathbf{p}_v)/d_{pv}$, $\boldsymbol{\rho}_{vr}=(\mathbf{p}_v-\mathbf{p}_r)/d_{vr}$ are defined in a common reference system, as illustrated in Fig.~\ref{fig:fig_12}.


Assuming half-wavelength-spaced antennas on each AP, the antenna array response for an azimuth angle $\omega$ is given as~\cite{behdad2024multi}
\vspace{-1mm}
\vspace{-1.75mm}
\begingroup\makeatletter\def\f@size{9.3}\check@mathfonts
\begin{equation} \mathbf{a}(\omega)=\big[ 1, e^{-j\omega}, \ldots, e^{-j(M_{\mathrm t}-1)\omega}\big]^\mathrm{T}\!\in\mathbb{C}^{M_{\mathrm t}\times1}.
\vspace{-1.25mm}
\end{equation} 
\endgroup

\noindent Accordingly, the array response from the $p$th transmitting AP and the $r$th receiving AP to the $v$th target can be expressed by $\mathbf{h}_{pv}=\mathbf{a}(\omega_{p,r,v}^{\mathrm{t}})$ and $\mathbf{h}_{vr}=\mathbf{a}(\omega_{p,r,v}^{\mathrm{r}})$, respectively. Considering target location uncertainty and pointing a sensing beam for target $v$ to its approximate position with an AoD of $\hat{\omega}_{p,r,v}^{\mathrm{t}}$ yield
\vspace{-1mm}
\vspace{-1mm}
\vspace{-1.25mm}
\begingroup\makeatletter\def\f@size{9.2}\check@mathfonts
\begin{equation} 
\hat{\mathbf{h}}_{pv}=\frac{1}{\sqrt{M_{\mathrm t}}}\mathbf{a}(\hat{\omega}_{p,r,v}^{\mathrm{t}}).\label{eq:sensing_BF}
\end{equation}
\endgroup
\vspace{-1.25mm}
\vspace{-1mm}
\vspace{-1mm}
\vspace{-1mm}

\noindent Then, the received signal at receiving AP $r$ is formulated as\footnote{For ease of analysis, this study assumes that the clutter can be appropriately mitigated by using the existing clutter suppression techniques~\cite{Zeng2025Multi,Liu2021Adaptive}, which is a commonly adopted model in the literature~\cite{Li2024seamless,Dehkordi2023Beam,Zhang2025Target}.}
\vspace{-1mm}
\vspace{-1mm}
\begingroup\makeatletter\def\f@size{9.2}\check@mathfonts
\begin{align}
\mathbf{y}_r\!=&\!\sum_{p=1}^{N_{\mathrm{tx}}}\sum_{v=1}^{T_{\mathrm{g}}}\beta_{p,r,v}^{\phantom{\dagger}}{\mathbf{H}}_{prv}^{\phantom{\dagger}}\bigg(\underbrace{\!\sum_{v^{\prime}=1}^{T_{\mathrm{g}}}\!\eta _{pv^{\prime}}^{\frac{1}{2}} \hat {\mathbf{H}}_{pv^{\prime}}^{\dagger}\mathbf{x}_{v^{\prime}}\!+\!\!\sum_{q=1}^{K_{\mathrm{u}}}\eta _{pq^{\phantom{\prime}}}^{\frac{1}{2}}\!\hat {\mathbf{H}}_{pq}^{\dagger}\mathbf{x}_{q}}_{ \triangleq\,\underline{\mathbf{x}}_{p} }\!\!\bigg)\hspace{-0.3mm}\!+\!\mathbf{w}_r,\nonumber\\[-15pt]
\label{eq:sensing_received}
\end{align}
\endgroup
\vspace{-1mm}
\vspace{-1mm}
\vspace{-1mm}
\vspace{-1mm}
\vspace{-1mm}
\vspace{-1.5mm}


\begingroup\makeatletter\def\f@size{9.5}\check@mathfonts
\noindent where $\mathbf{H}_{prv}\triangleq\mathbf{h}_{vr}^{\phantom{T}}\mathbf{h}_{pv}^\mathrm{T}\otimes \boldsymbol{\Psi}_{p,r,v}$ denotes the sensing reflected channel. Here, {\small$\beta_{p,r,v}\triangleq\alpha_{p,r,v}\xi_{p,r,v}^{1/2}$}, where $\alpha_{p,r,v}\sim\mathcal{CN}(0,\sigma_{p,r,v}^{2})$ is the radar cross-section (RCS) of the $v$th target, and $\xi_{p,r,v}=\frac{\lambda_{c}^{2}}{(4\pi)^{3}d_{pv}^{2}d_{vr}^{2}}$. After collecting the received signals $\mathbf{y}_r$ forwarded by each receiving AP, the overall sensing signal can be concatenated into $\mathbf{y}\!=\!\big[\mathbf{y}_{1}^{\mathrm{T}},\cdots,\mathbf{y}_{N_{\mathrm{rx}}}^{\mathrm{T}}\big]^{\mathrm{T}}\!\!\in\mathbb{C}^{N_{\mathrm{rx}}M_{\mathrm{t}}MN\times1}$ at the CPU.
\endgroup



\section{Multi-Static Position Estimation}
In this section, a general ML target position estimation scheme is developed for multi-static sensing systems, where the search space is reduced by using a direct position estimation rather than relying on range, velocity and angle estimates for every reflected path.


\begingroup\makeatletter\def\f@size{9.5}\check@mathfonts
Define the concatenated target position vector and the path coefficient vector as $\mathbf{p}\triangleq\big[\mathbf{p}_{1}^{\mathrm{T}},\mathbf{p}_{2}^{\mathrm{T}},\cdots,\mathbf{p}_{T_{\mathrm{g}}}^{\mathrm{T}}\big]^{\mathrm{T}}\in \mathbb{R}^{2T_{\mathrm{g}}\times1}$ and $\boldsymbol{\beta}\triangleq\big[\boldsymbol{\beta}_{1}^{\mathrm{T}},\boldsymbol{\beta}_{2}^{\mathrm{T}},\cdots,\boldsymbol{\beta}_{N_{\mathrm{rx}}}^{\mathrm{T}}\big]^{\mathrm{T}}\in \mathbb{C}^{N_{\mathrm{tx}}N_{\mathrm{rx}}T_{\mathrm{g}}\times1}$, respectively, where $\boldsymbol{\beta}_{r}=\big[\beta_{1,r,1},\cdots,\beta_{N_{\mathrm{tx}},r,T_{\mathrm{g}}}\big]^{\mathrm{T}}$. Then, the likelihood function of the received signal in~\eqref{eq:sensing_received} can be expressed as
\vspace{-1mm}
\vspace{-0.5mm}
\begin{align}
\hspace {-0.5pc}f(\mathbf{y}|\mathbf{p},\boldsymbol{\beta})=\!\prod_{r=1}^{N_{\mathrm{rx}}} \exp \!\Bigg(\!\!-\frac{1}{\sigma_w^2}\bigg\|\mathbf{y}_r-\!\sum_{p=1}^{N_{\mathrm{tx}}}\sum_{v=1}^{T_{\mathrm{g}}}\beta_{p,r,v}^{\phantom{\dagger}}{\mathbf{H}}_{prv}^{\phantom{\dagger}}\underline{\mathbf{x}}_{p} \bigg\|^2\Bigg),\!\!
\end{align}
\endgroup
\vspace{-1mm}
\vspace{-1mm}
\vspace{-0.5mm}

\noindent and the corresponding log-likelihood function is given by
\vspace{-0.5mm}
\vspace{-1mm}
\begingroup\makeatletter\def\f@size{9.5}\check@mathfonts
\begin{align}
\ell(\mathbf{y}|\mathbf{p},\boldsymbol{\beta})=&-\frac{1}{\sigma_w^{2}}\sum_{r=1}^{N_{\mathrm{rx}}}\bigg\|\mathbf{y}_r-\sum_{p=1}^{N_{\mathrm{tx}}}\sum_{v=1}^{T_{\mathrm{g}}}\beta_{p,r,v}^{\phantom{\dagger}}{\mathbf{H}}_{prv}^{\phantom{\dagger}}\underline{\mathbf{x}}_{p}\bigg\|^{2}.
\label{eq:likelihood_original}
\end{align}
\endgroup
\vspace{-1mm}
\vspace{-1mm}
\vspace{-1mm}

\noindent It should be noted that the estimate of each complex path coefficient $\boldsymbol{\beta}_r$ depends solely on the received signal $\mathbf{y}_r$ at the $r$th receiving AP. Consequently, the maximization of~\eqref{eq:likelihood_original} regarding $\boldsymbol{\beta}$ can be achieved by maximizing $N_{\mathrm{rx}}$ sub-functions~\eqref{eq:likelihood_r} at each receiving AP $r$, as shown at the bottom of the page.

\begin{figure*}[!hb] 
\vspace{-3.5mm}
\xhrulefill {black}{1pt}
\vspace{-0.5mm}
  \centering
    \begin{small}
    \begin{equation}
        \tilde{\ell}(\mathbf{y}_{r}|\mathbf{p},\boldsymbol{\beta}_r)\!=\!-\bigg\|\mathbf{y}_r-\sum_{p=1}^{N_{\mathrm{tx}}}\!\sum_{v=1}^{T_{\mathrm{g}}}\!\beta_{p,r,v}^{\phantom{\dagger}}{\mathbf{H}}_{prv}^{\phantom{\dagger}}\underline{\mathbf{x}}_{p}\bigg\|^{2}\!\!\!=\!-\!\left\|\mathbf{y}_r\right\|^2\!+2 \Re\Bigg\{ \!\sum_{p=1}^{N_{\mathrm{tx}}}\!\sum_{v=1}^{T_{\mathrm{g}}} \beta_{p,r,v}^{*}\!\left(\underline{\mathbf{x}}_{p}^{\dagger}\,\! {\mathbf{H}}_{prv}^{\dagger} \mathbf{y}_r^{\phantom{\dagger}}\!\right)\!\!\Bigg\}-\sum_{p^{\phantom{\prime}}\!\!=1}^{N_{\mathrm{tx}}}\!\sum_{p^{\prime}=1}^{N_{\mathrm{tx}}}\!\sum_{v^{\phantom{\prime}}\!\!=1}^{T_{\mathrm{g}}}\!\sum_{v^{\prime}=1}^{T_{\mathrm{g}}} \!\beta_{p,r,v}^{*}\beta_{p^{\prime}\!,r,v^{\prime}}^{\phantom{*}}\!\!\left( \underline{\mathbf{x}}_{p}^{\dagger}\;\! {\mathbf{H}}_{prv^{\phantom{\prime}}}^{\dagger}\!\! {\mathbf{H}}_{p^{\prime}\!rv^{\prime}}^{\phantom{\dagger}} \underline{\mathbf{x}}_{p^{\prime}}^{\phantom{\dagger}}\! \right)\!\!.
        \label{eq:likelihood_r}
    \end{equation}
    \end{small}
\end{figure*}

Let $u=vN_{\mathrm{tx}} + p$, and define the signal correlation vector $\mathbf{b}_r$ and matrix $\mathbf{A}_r\in\mathbb{C}^{N_{\mathrm{tx}}T_{\mathrm{g}}\times N_{\mathrm{tx}}T_{\mathrm{g}}}$ with elements given by
\vspace{-1mm}
\vspace{-1mm}
\begingroup\makeatletter\def\f@size{9.5}\check@mathfonts
\begin{subequations}
\begin{align}
\mathbf{b}_r[u]&=\underline{\mathbf{x}}_{p}^{\dagger}\;\! {\mathbf{H}}_{prv}^{\dagger} \mathbf{y}_r^{\phantom{\dagger}},\label{eq:Ar}\\
\mathbf{A}_r[u,u^{\prime}]&=\underline{\mathbf{x}}_{p}^{\dagger}\;\! {\mathbf{H}}_{prv}^{\dagger} {\mathbf{H}}_{p^{\prime}rv^{\prime}}^{\phantom{\dagger}} \underline{\mathbf{x}}_{p^{\prime}}^{\phantom{\dagger}},\label{eq:br}
\end{align}
\end{subequations}
\endgroup
\vspace{-1mm}
\vspace{-1mm}
\vspace{-1mm}
\vspace{-1mm}
\vspace{-1mm}

\noindent respectively. Upon discarding the irrelevant first term in the right-hand-side of~\eqref{eq:likelihood_r}, the log-likelihood sub-function can be reformulated as
\vspace{-0.5mm}
\vspace{-1mm}
\begingroup\makeatletter\def\f@size{9.5}\check@mathfonts
\begin{align}
\tilde{\ell}(\mathbf{y}_r|\mathbf{p},\boldsymbol{\beta}_r)=2 \Re\{\boldsymbol{\beta}_r^{\dagger} \hspace {0.1pc}\mathbf{b}_r^{\phantom{\dagger}}\}-\boldsymbol{\beta}_r^{\dagger}\hspace {0.1pc} \mathbf{A}_r\hspace {0.1pc} \boldsymbol{\beta}_r.
\label{eq:likelihood_Ab}
\end{align}
\endgroup
\vspace{-1mm}
\vspace{-1mm}
\vspace{-1mm}
\vspace{-1mm}
\vspace{-1mm}

\noindent The maximization with respect to $\boldsymbol{\beta}_r$ is readily obtained as $\hat{\boldsymbol{\beta}}_r= \mathbf{A}_r^{-1}\mathbf{b}_r^{\phantom{\dagger}}$. Finally, by substituting $\hat{\boldsymbol{\beta}}_r$ into~\eqref{eq:likelihood_Ab}, the reduced log-likelihood function with respect to the target position $\mathbf{p}$ is obtained as
\vspace{-1mm}
\vspace{-1mm}
\vspace{-1mm}
\begingroup\makeatletter\def\f@size{9.5}\check@mathfonts
\begin{align}
\ell_{1}(\mathbf{y}|\mathbf{p})=\sum_{r=1}^{N_{\mathrm{rx}}}\mathbf{b}_r^{\dagger}\mathbf{A}_r^{-1}\mathbf{b}_r^{\phantom{\dagger}},
\label{eq:likelihood_reduced}
\end{align}
\endgroup
\vspace{-1mm}
\vspace{-1mm}
\vspace{-1mm}

\noindent and the estimate of the target position can be obtained by maximizing~\eqref{eq:likelihood_reduced} as $\hat{\mathbf{p}}=\underset{\mathbf{p}\in\mathbb{R}^{2T_{\mathrm{g}}}}{\arg\max}\,\ell_{1}(\mathbf{y}|\mathbf{p})$.\footnote{Estimating the target positions via the ML estimator requires knowledge of the targets' velocities, which can be achieved by a simple modification of the ML estimator to include a search over the velocity dimension, yielding $\{\hat{\mathbf{p}},\hat{\mathbf{v}}\}\!=\!{\arg\max_{\{\mathbf{p},\mathbf{v}\}}}\ell_{1}(\mathbf{y}|\mathbf{p},\mathbf{v}).$ Due to page limitations, this study focuses on target position estimation and assumes that the target velocities are known.}

\begin{lemma}\label{lem1}
Assume the targets are sufficiently separated and the block size $MN$ is large, the ML estimator can be further simplified by
\vspace{-1mm}
\vspace{-1mm}
\vspace{-1mm}
\vspace{-1.8mm}
\begingroup\makeatletter\def\f@size{9.3}\check@mathfonts
\begin{align}
\hat{\mathbf{p}}=\underset{\mathbf{p}\in\mathbb{R}^{2T_{\mathrm{g}}}}{\arg\max}\sum_{r=1}^{N_{\mathrm{rx}}}\sum_{p=1}^{N_{\mathrm{tx}}}\sum_{v=1}^{T_{\mathrm{g}}}\frac{\big|\mathbf{y}_r^{\dagger}{\mathbf{H}}_{prv}^{\phantom{\dagger}}\underline{\mathbf{x}}_{p}\big|^{2}}{\big\|{\mathbf{H}}_{prv}^{\phantom{\dagger}}\underline{\mathbf{x}}_{p}\big\|^{2}}.
\label{eq:ML_estimator}
\end{align}
\endgroup

\end{lemma}

\begin{IEEEproof}
  Since the ISAC symbols in $\underline{\mathbf{x}}_{p}$ are independent zero-mean random variables, the off-diagonal terms $A_{u,u^{\prime}}$ are negligible for $u\neq u^{\prime}$~\cite{Dehkordi2023Beam}. Substituting~\eqref{eq:Ar} and~\eqref{eq:br} into~\eqref{eq:likelihood_reduced} yields the desired result.
\end{IEEEproof}


Based on the ML estimator in~\eqref{eq:likelihood_reduced} or~\eqref{eq:ML_estimator}, the search is discretized within the hotspot area by defining discrete search grids with coordinates $\left\{x_{\min}, \ldots, x_{\max}\right\} \times\left\{y_{\min}, \ldots, y_{\max}\right\}$, step sizes $\delta x$ and $\delta y$ are chosen to achieve the desired accuracy. Interestingly,~\eqref{eq:likelihood_reduced} can be interpreted as a summation of $N_{\mathrm{rx}}$ radar maps computed within the hotspot area~\cite{Pucci2025Cooperative}. Further, by neglecting the correlation terms in the matrix $\mathbf{A}_r$, the radar maps of each reflected path can be separabled, as shown in~\eqref{eq:ML_estimator}.

\begin{remark}
Unlike traditional multi-static position estimation methods that operate in a search space of dimension $\Omega=4N_{\mathrm{tx}}N_{\mathrm{rx}}T_{\mathrm{g}}$ by estimating the parameters of each reflected path~\cite{Dehkordi2024Multistatic}, the proposed direct ML estimator requires only a $\tilde{\Omega}=2T_{\mathrm{g}}$-dimensional search space, which significantly reduces computational complexity and enables real-time estimation. Optionally, the path parameters $\boldsymbol{\theta}_{p,r,v}^{(1)}$ can be calculated by~\eqref{eq:path_parameters} after obtaining the position estimate $\hat{\mathbf{p}}$.
\end{remark}


\section{Performance Metrics}
This section derives performance metrics for C\&S to evaluate the proposed ML estimation scheme and support subsequent system performance optimization. A closed-form CRLB and its low-complexity approximation are established for multi-static positioning error, which provide a theoretical lower bound on the mean square error (MSE) achievable by unbiased estimators for deterministic parameters.
\vspace{-1mm}

\subsection{The Original CRLB Expression}
To decouple the channel gain from the other geometric channel parameters, this study represents the channel parameters from the $p$th transmitting AP via the $v$th target to the $r$th receiving AP as
\vspace{-1mm}
\begingroup\makeatletter\def\f@size{9.2}\check@mathfonts
\begin{align}
\boldsymbol{\theta}_{p,r,v}\triangleq\big[\big(\boldsymbol{\theta}_{p,r,v}^{(1)}\big)^\mathrm{T},\big(\boldsymbol{\theta}_{p,r,v}^{(2)}\big)^\mathrm{T}\big]^\mathrm{T}\in\mathbb{R}^{6\times1},
\label{eq:theta}
\end{align}

\noindent where $\boldsymbol{\theta}_{p,r,v}^{(2)}\triangleq[\hspace{0.4mm}\beta_{p,r,v}^{(\mathrm{R})},\hspace{0.1mm}\beta_{p,r,v}^{(\mathrm{I})}\hspace{0.1mm}]^\mathrm{T}$ with $\beta_{p,r,v}^{(\mathrm{R})}=\Re\{\beta_{p,r,v}\}$ and $\beta_{p,r,v}^{(\mathrm{I})}=\Im\{\beta_{p,r,v}\} \vphantom{\underline{\beta_{p,r,v}^{(\mathrm{I})}}} $. Let $\bar{\mathbf{y}}_{r}[k,l]$ denote the noiseless part of the received signal at DD grid $[k,l]$ in~\eqref{eq:sensing_received}, then the $(i,j)th$ element of the Fisher information matrix (FIM) concerning $\boldsymbol{\theta}_{p,r,v}$ can be computed by~\cite{Pucci2025Cooperative}
\vspace{-1mm}
\vspace{-0.5mm}
\begin{equation}
\hspace {-0.5pc}[\mathbf{F}_{\boldsymbol{\theta}_{p,r,v}}]_{i,j}=\frac{2}{\sigma_w^2}\Re\left\{\sum_{k=0}^{N-1}\sum_{l=0}^{M-1}\!\left(\frac{\partial\bar{\mathbf{y}}_{r}[k,l]}{\partial\boldsymbol{\theta}_{p,r,v}[i]}\right)^{\!\dagger}\!\left(\frac{\partial\bar{\mathbf{y}}_{r}[k,l]}{\partial\boldsymbol{\theta}_{p,r,v}[j]}\right)\right\}\!,\!\!
\label{eq:FIM}
\end{equation}
\endgroup
\vspace{-1mm}
\vspace{-0.5mm}


\noindent which has a form of~\cite{AbuShaban2018Error}
\vspace{-1mm}
\begin{equation}
\hspace {-0.4pc}F_{x,x^{\prime}}\!=\!\Re\left\{(\text{RX~factor})\!\times\!(\text{TX~factor})\!\times\!(\text{signal factor})\right\}\!.\!
\label{eq:FIM_form}
\end{equation}
\vspace{-1mm}
\vspace{-1mm}
\vspace{-1mm}
\vspace{-1mm}
\vspace{-1mm}


\noindent For instance, it can be verified that
\vspace{-1mm}
\begingroup\makeatletter\def\f@size{9.5}\check@mathfonts
\begin{equation}
\hspace {-0.3pc}F_{\omega_{p,r,v}^{\mathrm{r}},\omega_{p,r,v}^{\mathrm{r}}}\!=\frac{2}{\sigma_w^2}\Re\{(\underbrace{ \beta_{p}^{*}\beta_{p}^{\phantom{*}}\dot{\mathbf{h}}_{vr}^{\dagger}\dot{\mathbf{h}}_{vr}^{\phantom{\dagger}} }_{\text{RX~factor}})\!\times\!( \underbrace{ \mathbf{h}_{pv}^{\dagger}\mathbf{V}_{p}^{\phantom{\dagger}}\mathbf{h}_{pv}^{\phantom{\dagger}} }_{\text{TX~factor}})\!\times\!\!\!\!\underbrace{R_{p,r,v}^{(0,0)}}_{\text{signal factor}}\!\!\!\}.\!
\label{eq:FIM_example}
\end{equation}
\endgroup
\vspace{-1mm}
\vspace{-1mm}

\noindent where $\dot{\mathbf{h}}_{vr}=\mathbf{c}_{M_{\mathrm{t}}}\!\odot\mathbf{h}_{vr}$, in which $\mathbf{c}_{M_{\mathrm{t}}}\!=[0,1,\ldots,M_{\mathrm{t}}-1]^\mathrm{T}$; and $\mathbf{V}_{p}\triangleq\sum_{v=1}^{T_\mathrm{g}}\eta_{pv}\hat{\mathbf{h}}_{pv}^{\phantom{\dagger}}\hat{\mathbf{h}}_{pv}^{\dagger} + \sum_{q_{\vphantom{1}}=1}^{K_\mathrm{u}}\sum_{i=1}^{L_{pq}^{\vphantom 1}}\eta_{pq}\hat{\mathbf{h}}_{pq,i}^{\phantom{\dagger}}\hat{\mathbf{h}}_{pq,i}^{\dagger} \vphantom{\underline{\hat{\mathbf{h}}_{pq,i}^{\phantom{\dagger}}\hat{\mathbf{h}}_{pq,i}^{\dagger}}} $. The remaining entries of~\eqref{eq:FIM} exhibit the structure in~\eqref{eq:FIM_form}, which are provided in Appendix~\ref{app:FIM_Entries}.

In adherence to~\eqref{eq:theta}, the FIM can be partitioned as
\vspace{-1mm}
\begingroup\makeatletter\def\f@size{9.5}\check@mathfonts
\begin{equation}
\mathbf{F}_{\boldsymbol{\theta}_{p,r,v}} = \left[\begin{matrix}
\mathbf{F}_{\boldsymbol{\theta}_{p,r,v}^{(1)}} & \mathbf{F}_{\boldsymbol{\theta}_{p,r,v}^{(1)}, \boldsymbol{\theta}_{p,r,v}^{(2)}}  \\
\mathbf{F}_{\boldsymbol{\theta}_{p,r,v}^{(1)}, \boldsymbol{\theta}_{p,r,v}^{(2)}}^\mathrm{T} & \mathbf{F}_{\boldsymbol{\theta}_{p,r,v}^{(2)}}  \\
\end{matrix}\right]. 
\end{equation}
\endgroup
\vspace{-1mm}

\noindent Consequently, the equivalent FIM of $\boldsymbol{\theta}_{p,r,v}^{(1)}$, which includes only the parameters related to the target position, is given by
\begingroup\makeatletter\def\f@size{9.5}\check@mathfonts
\begin{equation}
\mathbf{F}_{\boldsymbol{\theta}_{p,r,v}^{(1)}}^{\mathrm{e}}\!=\mathbf{F}_{\boldsymbol{\theta}_{p,r,v}^{(1)}}\!-\mathbf{F}_{\boldsymbol{\theta}_{p,r,v}^{(1)}, \boldsymbol{\theta}_{p,r,v}^{(2)}} \mathbf{F}_{\boldsymbol{\theta}_{p,r,v}^{(2)}}^{-1} \mathbf{F}_{\boldsymbol{\theta}_{p,r,v}^{(1)}, \boldsymbol{\theta}_{p,r,v}^{(2)}}^{\mathrm{T}}\!\in\mathbb{R}^{4\times4}.
\label{eq:EFIM}
\end{equation}
\endgroup

\noindent After collecting information from all reflected paths, the FIM for multi-static position estimation of target $v$ is given as
\vspace{-1mm}
\begingroup\makeatletter\def\f@size{9.5}\check@mathfonts
\begin{equation}
\mathbf{F}_{\mathbf{p}_v}=\sum_{p=1}^{N_{\mathrm{tx}}}\sum_{r=1}^{N_{\mathrm{rx}}}\nabla_{\mathbf{p}_v}^{\mathrm{T}}\boldsymbol{\theta}_{p,r,v}^{(1)}\, \mathbf{F}_{\boldsymbol{\theta}_{p,r,v}^{(1)}}^{\mathrm{e}}\!\!\nabla_{\mathbf{p}_v}^{\phantom{\mathrm{T}}}\boldsymbol{\theta}_{p,r,v}^{(1)}\in\mathbb{R}^{2\times2},
\label{eq:positionFIM}
\end{equation}
\endgroup
\vspace{-1mm}
\vspace{-1mm}
\vspace{-1mm}

\noindent where the Jacobian is expressed as
\renewcommand{\arraystretch}{1.25}
\begin{align}
\nabla_{\mathbf{p}_v} \boldsymbol{\theta}_{p,r,v}^{(1)}=&\left[\!\begin{array}{c}
\pi\, \mathbf{u}_r^{\mathrm{T}}\left( \frac{\mathbf{I}-\boldsymbol{\rho}_{vr} \boldsymbol{\rho}_{vr}^{\mathrm{T}}}{\left\| \mathbf{p}_v-\mathbf{p}_r \right\|} \right) \\
\pi\, \mathbf{u}_p^{\mathrm{T}}\left( \frac{\mathbf{I}-\boldsymbol{\rho}_{pv} \boldsymbol{\rho}_{pv}^{\mathrm{T}}}{\left\| \mathbf{p}_v-\mathbf{p}_p \right\|} \right) \\
\frac{1}{c}\left(\boldsymbol{\rho}_{vr}+\boldsymbol{\rho}_{pv}\right)\!\!^{\mathrm{T}}\\
\frac{\mathbf{v}_v^{\mathrm{T}}}{\lambda_c}\left( \frac{\mathbf{I}-\boldsymbol{\rho}_{vr} \boldsymbol{\rho}_{vr}^{\mathrm{T}}}{\left\| \mathbf{p}_v-\mathbf{p}_r \right\|} + \frac{\mathbf{I}-\boldsymbol{\rho}_{pv} \boldsymbol{\rho}_{pv}^{\mathrm{T}}}{\left\| \mathbf{p}_v-\mathbf{p}_p \right\|} \right) \\
\end{array}\!\right]
\!\triangleq\!\left[\!\begin{array}{c}
\boldsymbol{J}_{1}^{\mathrm{T}} \\
\boldsymbol{J}_{2}^{\mathrm{T}} \\
\boldsymbol{J}_{3}^{\mathrm{T}} \\
\boldsymbol{J}_{4}^{\mathrm{T}} \\
\end{array}\!\right].
\end{align}
\renewcommand{\arraystretch}{1}

Finally, the CRLB for the $v$th target's positioning error can be obtained as $\text{CRLB}_{\mathbf{p}_v}\!\!=\!\mathrm{Tr}\left( \mathbf{F}_{\mathbf{p}_v}^{-1} \right)$, and the PEB is defined as~\cite{Pucci2025Cooperative}
\begin{equation}
\text{PEB}_{\mathbf{p}_v}\triangleq\sqrt{\text{CRLB}_{\mathbf{p}_v}}.
\end{equation}

\subsection{Approximate Fisher Information}
Although the original FIM presents a closed-form expression for the calculation of the sensing CRLB, it is critical to note that its computational complexity scales as $\mathcal{O}\big(M^{2}N^{2}N_{\mathrm{tx}}N_{\mathrm{rx}}T_{\mathrm{g}}\big)$. In practice, the high dimensionality of matrix $\boldsymbol{\Psi}_{p,r,v}$ exacerbates this complexity, significantly hindering CRLB analysis and optimization. Therefore, developing a low-complexity expression becomes essential for practical implementation.


\begin{theorem}\label{thm1}
By considering only the beam directed toward the corresponding target, an approximation of the FIM for multi-static position estimation of target $v$ in~\eqref{eq:positionFIM} is given as
\vspace{-1mm}
\begin{equation}
\mathbf{F}_{\mathbf{p}_v}=\sum_{p=1}^{N_{\mathrm{tx}}}\sum_{r=1}^{N_{\mathrm{rx}}}\eta_{pv}\hat{\mathbf{F}}_{\mathbf{p}_{p,r,v}},
\label{eq:approxPositionFIM}
\vspace{-1mm}
\end{equation}
\noindent where $\hat{\mathbf{F}}_{\mathbf{p}_{p,r,v}}\!\approx\frac{2|\beta_{p}|^{2}}{\sigma_w^2}\big(d_{11}\boldsymbol{J}_{1}^{\phantom{\mathrm{T}}}\!\boldsymbol{J}_{1}^{\mathrm{T}}\!+d_{33}\boldsymbol{J}_{3}^{\phantom{\mathrm{T}}}\!\boldsymbol{J}_{3}^{\mathrm{T}}\!+d_{44}\boldsymbol{J}_{4}^{\phantom{\mathrm{T}}}\!\boldsymbol{J}_{4}^{\mathrm{T}}\big)$, with $d_{11}$, $d_{33}$ and $d_{44}$ given in~\eqref{eq:EFIM_derived}.

\end{theorem}

\begin{IEEEproof}
  The proof is given in Appendix~\ref{app:thm1}.
\end{IEEEproof} 

\begin{corollary}\label{cor1}
For the OFDM-signal-based systems, an approximation of the FIM for multi-static position estimation can be obtained by applying a similar procedure as in Theorem~\ref{thm1}, with the difference between them being that the signal factors are specified in~\eqref{eq:R_OFDM_derived}.
\end{corollary}

\begin{IEEEproof}
  The proof is given in Appendix~\ref{app:cor1}.
\end{IEEEproof}

\begin{remark}
  For sufficiently large $M$ and $N$ in practical implementation, the signal factors $R_{p,r,v}$ in~\eqref{eq:R_derived} and~\eqref{eq:R_OFDM_derived} are dominated by their first terms. Under this premise, the OTFS and OFDM-signal-based systems exhibit comparable CRLB under the same system bandwidth and time duration, aligning with existing studies~\cite{gaudio2020effectiveness}.
\end{remark}


\subsection{Communication Performance}
\vspace{-0.5mm}
In this study, communication SE $R_{q}\triangleq\omega_{x}\log _2\big( 1\!+\!\text{SINR}^{(\mathrm c)}_q \big)$, $x\in\{\text{otfs},\text{ofdm}\}$ is used as a performance metric, where the pre-log factors for the OTFS and OFDM-signal-based systems are given by $\omega_{\text{otfs}}=\frac{MN}{MN+N_{\mathrm{cp}}}$ and $\omega_{\text{ofdm}}=\frac{M}{M+N_{\mathrm{cp}}}$, respectively. Further, $\text{SINR}^{(\mathrm c)}_q$ is given by~\cite{Fan2024Power}
\vspace{-1mm}
\vspace{-1mm}
\begingroup\makeatletter\def\f@size{9.2}\check@mathfonts
\begin{equation}
    \hspace {-0.4pc}\text{SINR}^{(\mathrm c)}_q\!=\!\frac{\left( \sum_{p=1}^{N_{\mathrm{tx}}}{{\eta _{pq}^{1/2}}}b_{pq} \right)^2}{\sum_{p=1}^{N_{\mathrm{tx}}}\!\Big(\!\sum_{q^{\prime}=1}^{K_{\mathrm{u}}}{{\eta _{pq^{\prime}}}c_{pq,q^{\prime}}\!+\!\sum _{v=1}^{T_{\mathrm{g}}}{\eta _{pv}}c_{pq,v} }\Big)\!+\!\sigma_w^2},\!\label{eq:SINR_c}
\end{equation}
\endgroup
\vspace{-1mm}
\vspace{-1mm}

{\noindent where {\small$ b_{pq}\!\triangleq\!{\sum_{i=1}^{L_{pq}}}\mathrm{Tr}\left( \mathbf{B}_{pq,i}\right)$}, {\small$c_{pq,v}\!\triangleq\!\sum_{i=1}^{L_{pq}}{\mathrm{Tr}\left(\mathbf{B}_{pq,i}\mathbf{B}_{pv}\right)}$} and {\small$c_{pq,q^{\prime}}\!\triangleq\!\sum_{i=1}^{L_{pq}^{\phantom{1}}}\sum_{j=1}^{L_{pq^{\prime}}^{\phantom{1}}}{\mathrm{Tr}\left(\mathbf{B}_{pq,i}\mathbf{B}_{pq^{\prime}\!,j}\right)}$} are defined. Here, {\small$\mathbf{B}_{pq,i} \triangleq \mathbb{E} \{\hat{\mathbf{h}}_{pq,i}^{\phantom{\dagger}}\hat{\mathbf{h}}_{pq,i}^{\dagger}\}$} denotes the covariance matrix of estimated channel vector {\small$\hat{\mathbf{h}}_{pq,i}$} considering imperfect channel state information (CSI). Please refer to~\cite{Fan2024Power} for a detailed explanation of the channel estimation process. Meanwhile, by recalling~\eqref{eq:sensing_BF}, the sensing precoding matrix and its trace can be defined as {\small$\mathbf{B}_{pv}\triangleq\hat{\mathbf{h}}_{pv}^{\phantom{\dagger}}\hat{\mathbf{h}}_{pv}^{\dagger}$} and {\small$b_{pv}\!\triangleq\!\mathrm{Tr}\left( \mathbf{B}_{pv}\right)$}.

\section{Problem Formulation and Proposed Solutions}
\label{sec:Problem Formulation and Proposed Solutions}

\subsection{Joint AP Mode Selection and Power Allocation Design}
In this subsection, the joint AP mode selection and power allocation problem is formulated and solved. For ease of notation, we introduce the sets $\mathcal{N}\triangleq\{1,\ldots,N_{\mathrm{AP}}\}$, $\mathcal{Q}\triangleq\{1,...,K_{\mathrm u}\}$, and $\mathcal{T}\triangleq\{1,\ldots,T_{\mathrm{g}}\}$ to represent the collection of indices of the APs, users, and targets in the scenario, respectively. Define $\mathbf{a}$ as the binary vector that indicates the AP operation mode, with its $p$th element given by
\vspace{-2mm}
\begingroup\makeatletter\def\f@size{9.5}\check@mathfonts
\begin{equation}
a_{p}=\begin{cases}&\text{\!\!1,\; if AP }p\text{ operates in transmitting mode}\\
&\text{\!\!0,\; if AP }p\text{ operates in receiving mode,}\end{cases}
\end{equation}
\vspace{-2mm}

\noindent and $\boldsymbol{\eta}\in\mathbb{R}^{N_{\mathrm{AP}}(K_{\mathrm{u}}+T_{\mathrm{g}})\times1}$ is the concatenated power allocation coefficient vector, the max-min fairness optimization problem can then be expressed as follows:
\endgroup
\vspace{-1mm}
\vspace{-0.5mm}
\begingroup\makeatletter\def\f@size{9.2}\check@mathfonts
\begin{subequations}\label{eq:P1}
\begin{alignat}{2}
\textbf{P1:}~~&\underset{\mathbf{a},\boldsymbol{\eta}\geq\mathbf{0}}{\text{max}} &\quad& \underset{q\in\mathcal{Q}}{\text{min}}\Big\{\text{SINR}^{(\mathrm c)}_{q}\Big\} \label{eq:P1a}
\\ 
&~~ \text{s.t.} && \text{CRLB}_{\mathbf{p}_v}\!=\mathrm{Tr}\left( \mathbf{F}_{\mathbf{p}_v}^{-1} \right) \leq \gamma_{\mathrm s},\quad \forall v\in\mathcal{T}, \label{eq:P1b}
\\
& && P_{p}\leq a_{p}P_{\mathrm{d}},\quad \forall p\in\mathcal{N}, \label{eq:P1c}
\\
& && a_{p}\in \{0,1\},\quad \forall p\in\mathcal{N}, \label{eq:P1d}
\end{alignat}
\end{subequations}
\endgroup
\vspace{-0.5mm}
\vspace{-5mm}

\noindent where $\gamma_{\mathrm s}$ denotes the maximum sensing CRLB threshold, $\mathbf{F}_{\mathbf{p}_v}$ is modified to include $a_{p}$ using Theorem~\ref{thm1}, yielding
\vspace{-1mm}
\begingroup\makeatletter\def\f@size{9.3}\check@mathfonts
\begin{align}
\mathbf{F}_{\mathbf{p}_v}=\sum\nolimits_{p\in\mathcal{N}}\sum\nolimits_{p^{\prime}\in\mathcal{N}}a_{p}(1-a_{p^{\prime}})\eta_{pv}^{\phantom{*}}\hat{\mathbf{F}}_{\mathbf{p}_{p,p^{\prime}\!,v}},
\label{eq:Fisher_pt}
\end{align}
\endgroup
\vspace{-1mm}
\vspace{-1mm}
\vspace{-1mm}
\vspace{-1mm}

\noindent and~\eqref{eq:P1c} is the power constraint with $P_{p}$ given as follows.

\begin{lemma}\label{lem2}
The power constraint in~\eqref{eq:power_constraint} at the $p$th transmitting AP is given by
\vspace{-1mm}
\vspace{-0.5mm}
\begingroup\makeatletter\def\f@size{9.5}\check@mathfonts
\begin{equation}\label{eq:power_constraint_new}
P_{p}=\sum\nolimits _{q\in\mathcal{Q}}\eta _{pq}b_{pq}+\sum\nolimits _{v\in\mathcal{T}}\eta _{pv}b_{pv}.
\end{equation}
\endgroup
\end{lemma}

\vspace{-0.5mm}

\begin{IEEEproof}
  The proof is similar to that in~\cite[Appendix C]{Fan2024Power}, and is therefore omitted due to space limitation.
\end{IEEEproof}

To make this challenging mixed-integer non-convex problem tractable, we introduce an auxiliary variable $z\!=\!\text{min}_{q\in\mathcal{Q}}\,\text{SINR}^{(\mathrm c)}_q$ and relax the binary constraint~\eqref{eq:P1d} to $0\leq a_p \leq 1$ by noting that $a_p\in \{0,1\}$ is equivalent to $a_p=a_p^2$ under $0\leq a_p \leq 1$. The binary nature of $a_p$ at the optimal point is ensured by introducing a penalty term into the objective function to enforce $a_p=a_p^2$, where the penalty factor $\lambda$ indicates the relative importance of recovering binary values for $\mathbf{a}$ over fairness maximization~\cite{Elfiatoure2025Multiple}. For faster convergence, $a_p$ is replaced with $a_p^2$ in~\eqref{eq:P1c} in the optimization process. Accordingly, by applying the SCA method and using the following inequality
\vspace{-1mm}
\vspace{-0.5mm}
\begin{equation}~\label{eq:x2}
x^{2}\geq x_{0}(2x-x_{0}),
\end{equation}
\vspace{-1mm}
\vspace{-1mm}
\vspace{-1mm}
\vspace{-1mm}
\vspace{-1mm}
\vspace{-0.5mm}

\noindent the original optimization problem~\eqref{eq:P1} can be reformulated as
\vspace{-1mm}
\begingroup\makeatletter\def\f@size{9.2}\check@mathfonts
\begin{subequations}\label{eq:P2}
    \begin{align}
        \textbf{P2:}~~\underset{\mathbf{a}, \boldsymbol{\eta}, z}{\text{max}} \hspace{1em}& z-\lambda \sum\nolimits_{p\in\mathcal{N}}a_p-a_p^{(i)}\Big(2a_p-a_p^{(i)}\Big) \\
        \text{s.t.} \hspace{1em}\,  
        & \hspace{0.025cm}\text{SINR}\hspace{0.040cm}^{(\mathrm c)}_q \hspace{0.030cm}(\mathbf{a}, \boldsymbol{\eta})\geq \hspace{0.02cm}z,\ \;\hspace{0.02cm}q\in\mathcal{Q}, \label{eq:P2_SINR}\\
        & \text{CRLB}_{\mathbf{p}_v}(\mathbf{a}, \boldsymbol{\eta}) \leq\gamma_{\mathrm s},~v\in\mathcal{T}, \label{eq:P2_CRLB}\\
        &\!\sum\nolimits _{q\in\mathcal{Q}}\eta _{pq}b_{pq}+\!\sum\nolimits _{v\in\mathcal{T}}\eta _{pv}b_{pv}\leq a_p^{(i)}\Big(2a_p-a_p^{(i)}\Big)P_{\mathrm{d}},\nonumber\\
        &\qquad\quad\qquad p\in\mathcal{N},\\
        & 0\leq a_p \leq 1,~p\in\mathcal{N}. \label{eq:ap}
    \end{align}
\end{subequations}
\endgroup
\vspace{-1mm}
\vspace{-1mm}
\vspace{-1mm}
\vspace{-1mm}
\vspace{-1mm}

\addtocounter{equation}{7}
\begin{figure*}[hb] 
\vspace{-2.5mm}
\xhrulefill {black}{1pt}
\vspace{-0.5mm}
  \centering
  \begin{small}
    \begin{subequations}\label{eq:P3}
    \begin{align}
        \textbf{P3:}~~~\underset{\mathbf{a}, \boldsymbol{\eta}, z}{\text{max}} \hspace{1em}& z-\lambda \sum\nolimits_{p\in\mathcal{N}}a_p-a_p^{(i)}\Big(2a_p-a_p^{(i)}\Big) \\
        \mathrm{s.t.} \hspace{1em}\,  
        &f_{q}^{(i)}\Big(4\sum\nolimits_{p\in\mathcal{N}}{{\!\sqrt{a_{p}\eta _{pq}}}}b_{pq}\!-\!f_{q}^{(i)} z \Big)\!+\!\sum\nolimits_{p\in\mathcal{N}} \!\big(a_{p}^{(i)}\!-\!\mu_{pq}^{(i)}\big)\Big(2\big(a_{p}\!-\!\mu_{pq}\big)-\big(a_{p}^{(i)}\!-\!\mu_{pq}^{(i)}\big)\Big)\geq\sum\nolimits_{p\in\mathcal{N}} (a_{p}\!+\!\mu_{pq})^2 + 4\sigma_w^2,~\forall q\in\mathcal{Q},\!\\
        & \mathrm{Tr}\left( \left(\sum\nolimits_{p\in\mathcal{N}}\sum\nolimits_{{p^{\prime}}\in\mathcal{N}}\varrho_{p{p^{\prime}}v}\hat{\mathbf{F}}_{\mathbf{p}_{p,{p^{\prime}}\!,v}}\right)^{-1} \right) \leq \gamma_{\mathrm s},~\forall v\in\mathcal{T},\label{eq:P3_CRLB}\\
        &\sum\nolimits_{q\in\mathcal{Q}}\eta _{pq}b_{pq}+\sum\nolimits_{v\in\mathcal{T}}\eta _{pv}b_{pv}\leq a_p^{(i)}\Big(2a_p-a_p^{(i)}\Big)P_{\mathrm{d}},~ \forall p\in\mathcal{N},\\
        & 0\leq a_p \leq 1,~\forall p\in\mathcal{N}.
    \end{align}
    \end{subequations}
  \end{small}
\end{figure*}
\addtocounter{equation}{-8}

The non-convex nature of constraints~\eqref{eq:P2_SINR} and~\eqref{eq:P2_CRLB} make the resulting problem non-convex. To tackle the non-convexity of~\eqref{eq:P2_SINR}, we first rewrite the constraint as
\vspace{-1mm}
\vspace{-0.5mm}
\begingroup\makeatletter\def\f@size{9}\check@mathfonts
\begin{align}
&\hspace{-0.5pc}\frac{\left( \sum_{p\in\mathcal{N}}{{\sqrt{a_{p}\eta _{pq}}}}b_{pq} \right)^2}{z} \geq \nonumber \\
&\hspace{-0.5pc}\sum\nolimits_{p\in\mathcal{N}}a_{p}\Big(\sum\nolimits_{q^{\prime}\in\mathcal{Q}}{{\eta _{pq^{\prime}}}c_{pq,q^{\prime}}+\!\sum\nolimits_{v\in\mathcal{T}}{\eta _{pv}}c_{pq,v} }\Big)+\sigma_w^2.\!
\label{eq:P2_SINR_re}
\end{align}
\endgroup
\vspace{-1mm}
\vspace{-1mm}
\vspace{-1mm}
\vspace{-1mm}


\noindent For notational simplicity, let us define $\mu_{pq}\triangleq\sum\nolimits_{q^{\prime}\in\mathcal{Q}}{\eta _{pq^{\prime}}}c_{pq,q^{\prime}}+\sum\nolimits_{v\in\mathcal{T}}{\eta _{pv}}c_{pq,v}$, and~\eqref{eq:P2_SINR_re} can be equivalently reformulated as
\vspace{-1mm}
\begingroup\makeatletter\def\f@size{9}\check@mathfonts
\begin{align}
&\hspace{-0.5pc}\frac{\left( \!2\!\sum\nolimits_{p\in\mathcal{N}}{{\!\sqrt{a_{p}\eta _{pq}}}}b_{pq} \!\right)^2}{z}\!+\!\sum\nolimits_{p\in\mathcal{N}}\!\left( a_{p}\!-\!\mu_{pq} \right)^2\!\nonumber \\ 
&\qquad\qquad\qquad\qquad\qquad\geq\!\sum\nolimits_{p\in\mathcal{N}}\!\left( a_{p}\!+\!\mu_{pq} \right)^2\!+\!4\sigma_w^2.\!
\label{eq:P2_SINR_reformulated}
\end{align}
\endgroup
\vspace{-1mm}
\vspace{-1mm}
\vspace{-1mm}
\vspace{-1mm}

\noindent To handle the fractional term in the left-hand-side of the above inequality, we employ the following concave lower bound~\cite{Elfiatoure2025Multiple}
\vspace{-1mm}
\vspace{-1mm}
\vspace{-1mm}
\vspace{-1mm}
\vspace{-1mm}
\begingroup\makeatletter\def\f@size{9}\check@mathfonts
\begin{align}
\frac{x^{2}}{y}\geq
\frac{x_{0}}{y_{0}}\bigl(2x-\frac{x_{0}}{y_{0}}y\bigr),
\end{align}
\vspace{-1mm}
\vspace{-1mm}
\vspace{-1mm}
\vspace{-1mm}
\vspace{-0.5mm}

\noindent and define
\begin{equation}
f_{q}^{(i)}\triangleq\frac{ 2\sum_{p\in\mathcal{N}}{{\sqrt{a_{p}^{(i)}\eta _{pq}^{(i)}}}}b_{pq} }{z^{(i)}}.
\end{equation}
\endgroup
\vspace{-1mm}
\vspace{-1mm}
\vspace{-1mm}
\vspace{-1mm}
\vspace{-0.5mm}

\noindent Then, the final convex approximation of~\eqref{eq:P2_SINR_reformulated} is given by
\vspace{-1mm}
\begingroup\makeatletter\def\f@size{9}\check@mathfonts
\begin{align}
&f_{q}^{(i)}\Big(4\sum\nolimits_{p\in\mathcal{N}}{{\sqrt{a_{p}\eta _{pq}}}}b_{pq}\!-\! f_{q}^{(i)} z \Big)\nonumber\\
&+\!\sum\nolimits_{p\in\mathcal{N}}\! \big(a_{p}^{(i)}\!-\!\mu_{pq}^{(i)}\big)\Big(2\big(a_{p}-\mu_{pq}\big)-\big(a_{p}^{(i)}-\mu_{pq}^{(i)}\big)\Big)\nonumber\\
&\geq\sum\nolimits_{p\in\mathcal{N}} (a_{p}\!+\!\mu_{pq})^2 \!+\! 4\sigma_w^2,
\end{align}
\vspace{-1mm}
\vspace{-1mm}
\vspace{-1mm}
\vspace{-1mm}

\noindent where the inequality~\eqref{eq:x2} is used, with $x$ and $x_0$ replaced by $a_{p}-\mu_{pq}$ and $a_{p}^{(i)}-\mu_{pq}^{(i)}$, respectively.
\endgroup

\begingroup\makeatletter\def\f@size{9}\check@mathfonts
The focus now shifts to the sensing CRLB constraint~\eqref{eq:P2_CRLB}. Based on the SCA method, the bilinear and trilinear terms can be approximated by the first-order Taylor expansions at the iteration point $(a_p^{(i)},a_{p^{\prime}}^{(i)},\eta_{pv}^{(i)})$ as follows:
\vspace{-1mm}
\vspace{-0.5mm}
\begin{subequations}
\begin{align}
a_p \eta_{pv} &\approx a_p^{(i)} \eta_{pv}^{\phantom 1} + a_p^{\phantom 1} \eta_{pv}^{(i)} - a_p^{(i)} \eta_{pv}^{(i)}, \\
\hspace{-0.5pc}a_p a_{p^{\prime}} \eta_{pv} &\approx a_p^{(i)} a_{p^{\prime}}^{(i)} \eta_{pv}^{\phantom 1} \!+ \!\left( a_p^{(i)} a_{p^{\prime}}^{\phantom 1} \!+ a_p^{\phantom 1} a_{p^{\prime}}^{(i)} \!- 2 a_p^{(i)} a_{p^{\prime}}^{(i)} \right) \eta_{pv}^{(i)}.\!\!
\end{align}
\end{subequations}
\vspace{-1mm}
\vspace{-1mm}
\vspace{-1mm}
\vspace{-1mm}
\vspace{-1mm}

\noindent Therefore, the coefficient in~\eqref{eq:Fisher_pt} can be linearized as
\vspace{-1mm}
\vspace{-0.5mm}
\begin{align}
&\hspace {-0.5pc}a_{p}(1-a_{p^{\prime}})\eta_{pv}=a_{p}\eta_{pv}-a_{p}a_{p^{\prime}}\eta_{pv} \approx a_p^{(i)}\big(1 - a_{p^{\prime}}^{(i)}\big)\eta_{pv} \nonumber \\
&\hspace {-0.2pc} + \!\left(a_p\big(1 - a_{p^{\prime}}^{(i)}\big) + a_p^{(i)}\big(2a_{p^{\prime}}^{(i)} - a_{p^{\prime}} - 1\big)\right)\eta_{pv}^{(i)}\!\triangleq\varrho_{p{p^{\prime}}v}.\!
\end{align}
\vspace{-1mm}
\vspace{-1mm}
\vspace{-1mm}
\vspace{-1mm}
\vspace{-1mm}
\vspace{-0.5mm}
\endgroup

\begin{small}
\begin{algorithm}[t]
\caption{Proposed Joint AP Mode Selection and Power Allocation Algorithm for Problem $\textbf{P1}$}
\label{alg1}
\begin{algorithmic}[1]
\STATE $\textbf{Initialization:}$ Set the iteration counter $i\!=\!0$, the penalty factor $\lambda\!>\!1$, and an arbitrary feasible set $\mathbf{x}^{(0)}\!\triangleq\!\{\mathbf{a}^{(0)}\!, \boldsymbol{\eta}^{(0)}\}$.
\vspace{-4.5mm}
\REPEAT
\STATE $i\leftarrow i+1$.
    \STATE Update $\mathbf{x}^{(i)}$ by solving the convex optimization problem in~\eqref{eq:P3};
\UNTIL convergence.
\STATE $\textbf{Output:}$ The AP mode selection vector $\mathbf{a}^{(i)}$ and the transmit power coefficients $\boldsymbol{\eta}^{(i)}$.
\end{algorithmic}
\end{algorithm}
\end{small}

\noindent Finally, the convex optimization problem is obtained as~\eqref{eq:P3}, which is shown at the bottom of the page. The overall algorithm for solving the joint AP mode selection and power allocation problem is summarized in \textbf{Algorithm~\ref{alg1}}. The convergence analysis follows the proof of~\cite[Proposition 1]{Wang2024Bidirectional}, thus is omitted due to space constraints.
\stepcounter{equation}

\emph{Complexity of Algorithm 1:}~\textbf{Algorithm~\ref{alg1}} requires to solve a series of convex problems~\eqref{eq:P3}. Using the Schur complement, constraint~\eqref{eq:P3_CRLB} is equivalent to
\vspace{-1mm}
\begin{align}
\vspace{-0.5mm}
\begin{bmatrix}\hat{\mathbf{F}}_{\mathbf{p}_v}\!\!&\mathbf{I}_2\\\mathbf{I}_2&\mathbf{S}_v\end{bmatrix}\succeq0,\quad\mathrm{Tr}(\mathbf{S}_v)\leq\gamma_s,
\end{align}
\vspace{-1mm}
\vspace{-1mm}
\vspace{-1mm}
\vspace{-0.5mm}


\noindent where $\hat{\mathbf{F}}_{\mathbf{p}_v} \!=\! \sum_{p,p'} \varrho_{p{p^{\prime}}v}\hat{\mathbf{F}}_{\mathbf{p}_{p,{p^{\prime}}\!,v}}$. Then, Problem~\eqref{eq:P3} can be equivalently transformed to a semidefinite program that involves $A_v\triangleq N_{\mathrm{AP}}(2K_{\mathrm u}+T_\mathrm{g}+1)+3T_\mathrm{g}+1$ scalar variables. Therefore, the algorithm for solving Problem~\eqref{eq:P3} requires a complexity of $\mathcal{O}\left(A_v^{4.5}\log(1/\delta)\right)$ in each iteration, where $\delta> 0$ is the accuracy of the interior-point method~\cite{Ma2023Robust,Luo2010Semidefinite}. The convergence performance will be illustrated by the numerical results in Section~\ref{sec:Numerical Results}.

\vspace{-1mm}

\subsection{Low-Complexity Design with Closest AP Mode Selection}
This subsection introduces a low-complexity design to reduce the complexity of the joint optimization problem while ensuring acceptable system performance. To this end, the original optimization problem is decomposed into two disjoint sub-tasks: 1) AP mode selection and 2) AP power allocation. In the first stage, AP mode selection is performed based on the distance to the targets' hotspot area~\cite{Fan2024Power}. Next, a power allocation problem is solved to optimize the power allocation coefficients at the transmitting APs for the given AP modes.


\emph{1) Closest AP Mode Selection:}~~Let $\mathcal{N}_{\mathrm{Tx}}$ and $\mathcal{N}_{\mathrm{Rx}}$ denote the sets containing the indices of ISAC transmitting APs and sensing receiving APs, respectively. Our closest AP mode selection method is summarized in \textbf{Algorithm~\ref{alg2}}, where $\dot{d}_{vr}$ denotes the distance from AP $r$ to the $v$th target’s sensing hotspot area.

Initially, all APs operate in transmitting mode, with the exception of those closest to each target's hotspot area serving as sensing receivers. Next, during each iteration, the transmitting AP closest to the hotspot is switched to receiving mode, after which the power allocation scheme is employed to maximize the minimum SE among all users. This procedure continues until the minimum SE ceases to improve.

\begin{algorithm}[t]
\caption{Closest AP Mode Selection}
\label{alg2}
\begin{algorithmic}[1]
\STATE $\textbf{Initialization:}$ Set $\mathcal{N}_{\mathrm{Rx}}=\{{\operatorname*{\arg\min}}_{r\in\mathcal{N}}\,\dot{d}_{vr}, \forall v\in\mathcal{T}\}$ and $\mathcal{N}_{\mathrm{Tx}}=\mathcal{N}\setminus\mathcal{N}_{\mathrm{Rx}}$. Given the tolerance $\epsilon>0$, $i=0$.
\STATE Calculate $\mathrm{SE}^{(0)}\!=\text{min}_{q\in\mathcal{Q}}\,R_{q}\left(\mathcal{N}_{\mathrm{Tx}},\mathcal{N}_{\mathrm{Rx}}\right)$ based on the power allocation strategy outlined in Algorithm~\ref{alg3}.
\REPEAT
\STATE $i\leftarrow i+1$.
    \STATE Set $r^{*}={\operatorname*{\arg\min}}_{v\in\mathcal{T},r\in\mathcal{N}_{\mathrm{Tx}}}\dot{d}_{vr}$;
    \STATE Calculate $\mathrm{SE}^{(i)}=\text{min}_{q\in\mathcal{Q}}\,R_{q}(\mathcal{N}_{\mathrm{Tx}}\setminus r^{*},\mathcal{N}_{\mathrm{Rx}}\cup r^{*})$.
    \IF{$\mathrm{SE}^{(i)}\geq\mathrm{SE}^{(i-1)}$}
        \STATE Update $\mathcal{N}_{\mathrm{Tx}}=\mathcal{N}_{\mathrm{Tx}}\setminus r^{*}$ and $\mathcal{N}_{\mathrm{Rx}}=\{\mathcal{N}_{\mathrm{Rx}}\cup r^{*}\}$;
    \ENDIF
\UNTIL $\mathrm{SE}^{(i)}-\mathrm{SE}^{(i-1)}\leq\epsilon$.
\STATE $\textbf{Output:}$ $\mathcal{N}_{\mathrm{Tx}}$ and $\mathcal{N}_{\mathrm{Rx}}$, i.e., the indices of APs operating in transmitting and receiving mode, respectively.
\end{algorithmic}
\end{algorithm}

\emph{2) Power Allocation:}~~For a given AP mode selection vector $\mathbf{a}$, the original optimization problem~\eqref{eq:P1} is reduced to
\vspace{-1mm}
\begingroup\makeatletter\def\f@size{9}\check@mathfonts
\begin{subequations}\label{eq:35}
\begin{alignat}{2}
\textbf{P4:}~~&\underset{\boldsymbol{\eta}\geq\mathbf{0}}{\text{max}} &\quad& \underset{q\in\mathcal{Q}}{\text{min}}\Big\{\text{SINR}^{(\mathrm c)}_{q}\Big\} \label{eq:35a}
\\[-0.2mm]
&~ \text{s.t.} && \text{CRLB}_{\mathbf{p}_v} \leq \gamma_{\mathrm s},\quad v\in\mathcal{T}, \label{eq:35b}
\\[-0.3mm]
& && P_{p}\leq P_{\mathrm{d}},\quad\quad\quad\! p\in\mathcal{N}_{\mathrm{Tx}}. \label{eq:35c}
\end{alignat}
\end{subequations}
\endgroup
\vspace{-1mm}
\vspace{-1mm}
\vspace{-1mm}
\vspace{-1mm}

Since the trace of the inverse, {\small$\mathrm{Tr}\left( \mathbf{X}^{-1} \right)$}, is convex, it can be verified that the objective function in~\eqref{eq:SINR_c} exhibits a fractional programming structure, whereas the constraints \eqref{eq:35b}, \eqref{eq:35c} are convex. Therefore, by applying the quadratic transform~\cite{shen2018fractional}, the power allocation Problem~\eqref{eq:35} can be reformulated as a convex optimization problem, which can be expressed as
\vspace{-1mm}
\begingroup\makeatletter\def\f@size{9.2}\check@mathfonts
\begin{subequations}
\begin{alignat}{2}
\hspace {-0.4pc}\textbf{P5:}~~& \underset{\boldsymbol{\eta},z}{\text{max}} &\ \ \!& z 
\\[-0.6mm]
&~ \text{s.t.} && \mathrm{Tr}\left( \mathbf{F}_{\mathbf{p}_v}^{-1} \right) \leq \gamma_{\mathrm s},\quad v\in\mathcal{T},
\\[-0.15mm]
& && \sum\nolimits _{q\in\mathcal{Q}}\eta _{pq}b_{pq}+\!\sum\nolimits _{v\in\mathcal{T}}\eta _{pv}b_{pv}\leq P_{\mathrm{d}},\quad\hspace{-0.5mm} p\in\mathcal{N}_{\mathrm{Tx}},\!\!
\\[-0.15mm]
& && -y_q^2\sum\nolimits_{p\in\mathcal{N}_{\mathrm{Tx}}}\!\Big(\sum\nolimits_{q^{\prime}\in\mathcal{Q}}{\eta _{pq^{\prime}}}c_{pq,q^{\prime}}\!+\!\sum\nolimits_{v\in\mathcal{T}}{\eta _{pv}}c_{pq,v}\Big)\!\!\nonumber \\[-0.15mm]
& && \quad\ \!-y_q^2\sigma_w^2+2y_q\sum\nolimits_{p\in\mathcal{N}_{\mathrm{Tx}}}\!\eta _{pq}^{1/2}b_{pq}\geq z,\quad \!q\in\mathcal{Q},\!\!
\end{alignat}
\label{eq:opt_main}
\end{subequations}
\endgroup

\noindent where the auxiliary variable $y_q$ for fixed $\boldsymbol{\eta}$ is defined as
\vspace{-1mm}
\begin{equation} 
\hspace {-0.4pc}y_q=\frac{\sum _{p\in\mathcal{N}_{\mathrm{Tx}}}\eta_{pq}^{1/2}b_{pq}}{\sum _{p\in\mathcal{N}_{\mathrm{Tx}}}\!\Big(\!\sum _{q^{\prime}\in\mathcal{Q}}\eta_{pq^{\prime}}c_{pq,q^{\prime}}\!+\!\sum_{v\in\mathcal{T}}{\eta _{pv}}c_{pq,v}\Big)\!+\!\sigma_w^2}.\!\!
\label{eq:opt_yq}
\end{equation}
\vspace{-1mm}
\vspace{-1mm}

\noindent The optimization problem can be solved through an iterative approach, as outlined in \textbf{Algorithm~\ref{alg3}}. Similarly, Problem~\eqref{eq:opt_main} can be equivalently transformed to a semidefinite program that involves $\dot{A}_v\triangleq N_{\mathrm{tx}}(K_{\mathrm u}+T_\mathrm{g})+3T_\mathrm{g}+1$  scalar variables. Therefore, the algorithm for solving Problem~\eqref{eq:opt_main} requires a complexity of $\mathcal{O}\big(\dot{A}_v^{4.5}\log(1/\delta)\big)$ in each iteration, where $\delta> 0$ is the accuracy of the interior-point method~\cite{Ma2023Robust,Luo2010Semidefinite}.

\begin{small}
\begin{algorithm}[t]
\caption{Power Allocation with Fixed AP Modes}
\label{alg3}
\begin{algorithmic}[1]
\STATE $\textbf{Initialization:}$ Set an arbitrary initial positive $\boldsymbol{\eta}^{(0)}$, the tolerance $\epsilon>0$ and the maximum iteration number $I$. Set the iteration counter to $i=0$ and $z^{(0)}=0$.
\REPEAT
\STATE $i\leftarrow i+1$.
    \STATE Update $y_q^{(i)}$ according to \eqref{eq:opt_yq};
    \STATE Update $\boldsymbol{\eta}^{(i)}$ by solving the convex optimization problem \eqref{eq:opt_main} for fixed $y_q$;
\UNTIL $|z^{(i)}-z^{(i-1)}|\leq\epsilon$ \OR $i=I$.
\STATE $\textbf{Output:}$ The transmit power coefficients $\boldsymbol{\eta}^{(i)}$.
\end{algorithmic}
\end{algorithm}
\end{small}

\begin{table}[!t]
\vspace{3mm}
\centering
\caption{Simulation Parameters}
\label{tab2}
\rowcolors{2}{lightgray!30}{white}
\renewcommand\arraystretch{1.45}{
    \centering
    \begin{tabular}{ccc}
    \Xhline{0.8pt}\rowcolor{gray!35}
        \textbf{Parameters} & \textbf{Symbol} & \textbf{Value} \\ \Xhline{0.5pt}
        Carrier frequency & $f_{c}$ & 38\,GHz \\ 
        Bandwidth & $B$ & 64\,MHz \\ 
        Number of subcarriers & $M$ & 128 \\  
        Number of symbols & $N$ & 128 \\  
        Scenario size & - & 300m\,$\times$\,300m \\   
        Number of APs & $N_{\mathrm{AP}}$ & 32 \\  
        Number of antennas at each AP & $M_{\mathrm t}$ & 16 \\  
        Number of users & $K_{\mathrm u}$ & 10 \\ 
        Number of targets & $T_{\mathrm{g}}$ & 2 \\ 
        Maximum speed $($UE$/$Target$)$ & $v_{\max}$ & 300\,km/h \\  
        CP sample length & $N_{\mathrm{cp}}$ & $\lceil\tau_{\max}M\Delta f\rceil$ \\ 
        Sensing PEB threshold & $\gamma_{\mathrm s}$ & 0.1\,m \\ 
        Maximum transmit power & $P_{\mathrm{d}}$ & 1\,W \\ 
        Noise variance & $\sigma_w^2$ & -89\,dBm \\  
        RCS variance & $\sigma_{rcs}^2$ & 0\,dBsm \\ \Xhline{0.8pt}
    \end{tabular}}
    \vspace{-18pt}
\end{table}

\section{Numerical Results}
\label{sec:Numerical Results}
This section presents a comprehensive numerical analysis to evaluate the performance of applying the OTFS signal to the CF-ISAC systems. The key simulation parameters are listed in Table~\ref{tab2} unless otherwise specified. The path loss for the communication and sensing channels is modeled by the 3GPP Urban Microcell model and radar equation, respectively. A total area of 300\,m\,$\times$\,300\,m is considered within which $N_{\mathrm{AP}}$ APs, $K_{\mathrm u}$ users, and $T_{\mathrm{g}}$ targets are randomly distributed. The users and targets are moving at a maximum speed of 300\,km/h, and the AP antenna arrays $\mathbf{u}_p$ and $\mathbf{u}_r$ are randomly directed. The maximum downlink transmit power is set to 1\,W and 5\,W for CF APs and cellular AP, respectively. The RCS is modulated by the Swerling-I model, where $\alpha_{p,r,v}\hspace{-0.4mm}\sim\hspace{-0.2mm}\mathcal{CN}(0,\sigma_{p,r,v}^{2})$ remains constant during the sensing period~\cite{behdad2024multi}. For simplicity, it is assumed that the RCS values are independent and have the same variance $\sigma_{p,r,v}^2\hspace{-0.3mm}=\sigma_{rcs}^2$ for each reflected path. The channel estimation for the OTFS and OFDM-signal-based systems employs embedded pilot (EP) and block-based (BT) methods, respectively~\cite{Fan2024Power}.

To evaluate the efficiency of the proposed AP mode selection schemes, we introduce a random AP mode benchmark. Therefore, this section compares three distinct algorithms: i)~the joint AP mode selection and power allocation algorithm (\emph{JAP}); ii) the closest AP selection and power allocation algorithm (\emph{CAP}); and iii) the random AP selection and power allocation algorithm (\emph{RAP}). For a fair comparison, the number of receiving APs for the \emph{RAP} algorithm, $N_{\mathrm{rx}}$, is set to be close to the average optimized values obtained from the proposed \emph{JAP} scheme.

\begin{figure}[!t]
    \centering
    \includegraphics[width=2.95 in]{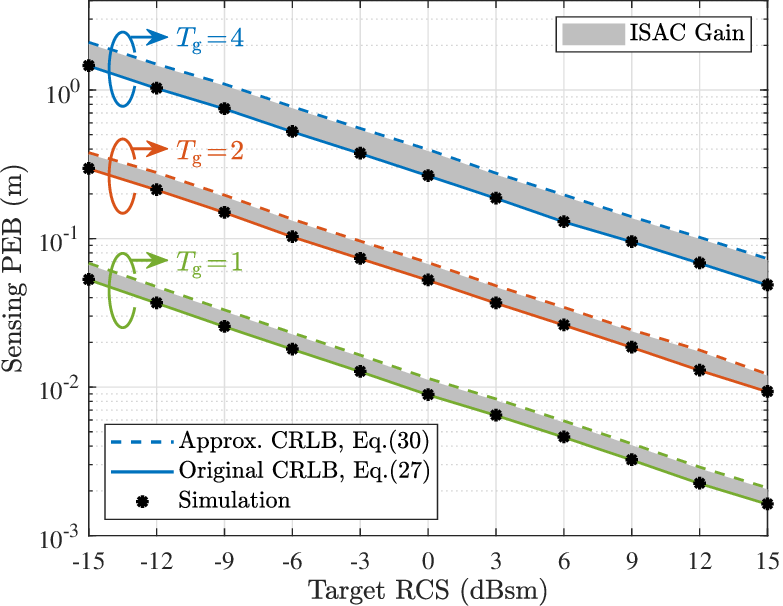}
    \vspace{-0.2cm}
    \caption{The sensing PEB versus different target numbers and RCS variances.}
    \label{fig:fig_2}
    \vspace{-0.2cm}
\end{figure}

\subsection{Verification of the PEB Expressions}
Fig.~\ref{fig:fig_2} presents the analytical sensing PEB calculated by~\eqref{eq:positionFIM} and the approximation in~\eqref{eq:approxPositionFIM}, along with the corresponding simulated results. Due to the high computational complexity of the original expression, we employ the closest AP mode selection method where transmit power optimization is not considered. Instead, the APs transmit with equal power, and the power allocation coefficients at the $p$th transmitting AP are set to $\eta_{pq}\hspace{-0.3mm}=\hspace{-0.3mm}\eta_{pv}\hspace{-0.3mm}=\hspace{-0.3mm}P_{\mathrm{d}}/\big(\sum _{q\in\mathcal{Q}}b_{pq}+\!\sum _{v\in\mathcal{T}}b_{pv}\big)$. Since the effects of the communication beams and the sensing beams directed at other targets are not considered, the proposed approximation serves as an upper bound of the original PEB, with their gap representing the ISAC coordination gain. Further, it is observed that the positioning accuracy degrades as the number of simultaneously sensed targets increases, in which case ISAC becomes more beneficial with an enhanced coordination gain.

\begin{figure}[!t]
    \vspace{-0.1cm}
    \centering
    \includegraphics[width=2.95 in]{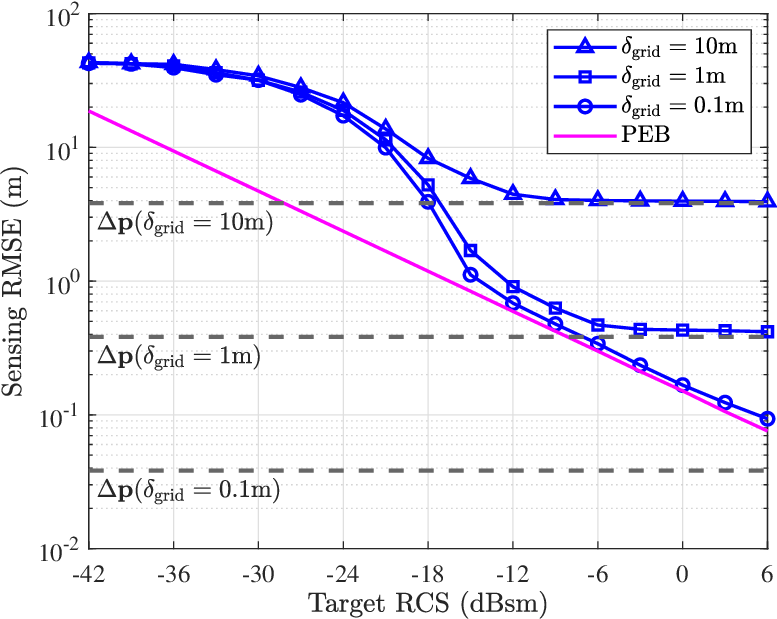}
    \vspace{-0.2cm}
    \caption{The sensing RMSE and PEB performance versus different ML estimator steps and RCS variances ($N_{\mathrm{AP}}=32, T_{\mathrm{g}}=1, M_{\mathrm t}=16$).}
    \label{fig:fig_4}
    \vspace{-0.25cm}
\end{figure}

\subsection{ML Estimation Performance}

To validate the effectiveness of the proposed ML position estimation scheme, we compare its estimate root MSEs (RMSEs) against the corresponding PEBs for various target RCS values. For a given target position $\mathbf{p}$ and its estimate $\hat{\mathbf{p}}$, the position estimate RMSE is computed as $\text{RMSE}=\sqrt{\frac{1}{N} \sum_{n=1}^{N}\left\|\hat{\mathbf{p}}_n-\mathbf{p}\right\|^2}$, where $N$ is the number of Monte Carlo iterations~\cite{Pucci2025Cooperative}. The ML search uses square grids with step sizes of 10~m, 1~m, and 0.1~m. As shown in Fig.~\ref{fig:fig_4}, an increase in the target RCS improves the sensing SNR, reducing both RMSE and PEB. In addition, the RMSE performance is also constrained by the step of the ML search grid $\delta_{\text{grid}}$. Due to the random location of the target relative to the discrete grid points, the RMSE asymptotically converges to the resolution of the ML search grid as the target RCS increases. For square grids, this resolution is given by $\Delta\mathbf{p}=\delta_{\text{grid}}/\sqrt{6}$. Further, the results demonstrate that with a sufficiently high target RCS (e.g., above -12~dB) and fine search grids ($\delta_{\text{grid}} = 0.1$~m), the proposed ML estimator achieves promising performance closely approaches the PEB, highlighting the high accuracy of the proposed method.

\begin{figure}[!t]
    \centering
    \includegraphics[width=2.95 in]{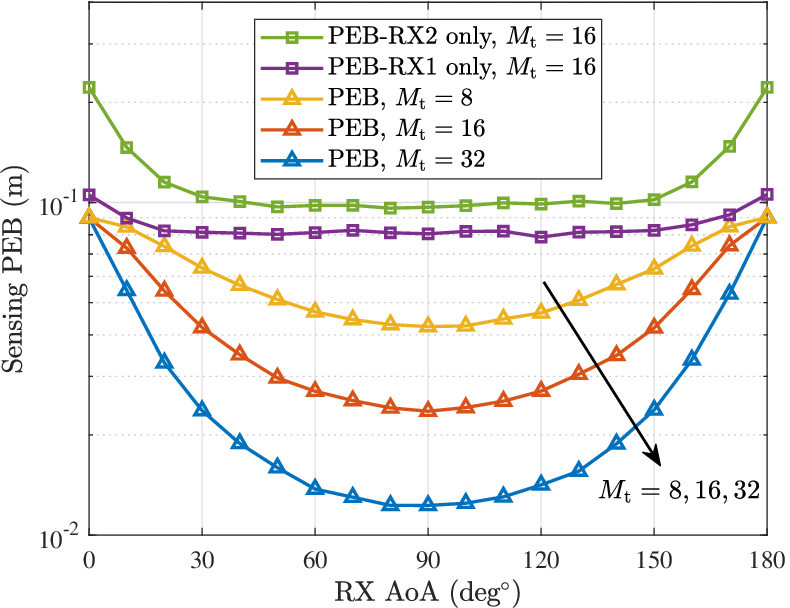}
    \vspace{-0.2cm}
    \caption{The sensing PEB versus different receiving AP array antennas and AoAs ($N_{\mathrm{AP}}=32, T_{\mathrm{g}}=1, \sigma_{rcs}^2=0$ dBsm).}
    \label{fig:fig_8}
    \vspace{-0.28cm}
\end{figure}

\subsection{Impact of Receiving AP Locations and AoAs}
Fig.~\ref{fig:fig_8} shows the impact of the AoAs on the PEB, including individual PEB contributions from the closer receiver RX1 and the farther receiver RX2. Specifically, when the AoA is close to 0° or 180°, the antenna element vector~$\mathbf{u}_r$ becomes parallel to $\boldsymbol{\rho}_{vr}\!$ (see Fig.~\ref{fig:fig_12}), leading to infinitely large angular estimation error. Consequently, increasing the number of antennas provides no improvement to estimation accuracy. Conversely, an AoA of 90° minimizes the angular estimation error, which reveals the vital role of the AP antenna array directions in improving estimation performance.\footnote{To address the challenge of simultaneously pointing a single receive antenna array towards multiple targets, digital receive beamforming emerges as a promising focus for future research.} As expected, the PEB using solely RX2 is higher than that when only RX1 is used due to its greater distance from the target. In the single-receiver case, the PEB curve exhibits a flat bottom shape where adjustments to the array direction yield no further improvement. This is because the overall PEB is constrained by a combination of range, Doppler, and angular estimation errors, while the angular estimation error is no longer the dominant error source in this section of the curve. In contrast, the collaborative use of both RX1 and RX2 significantly reduces the PEB, demonstrating the benefits of multi-static sensing.

\begin{figure}[!t]
    \centering
    \includegraphics[width=3.2 in]{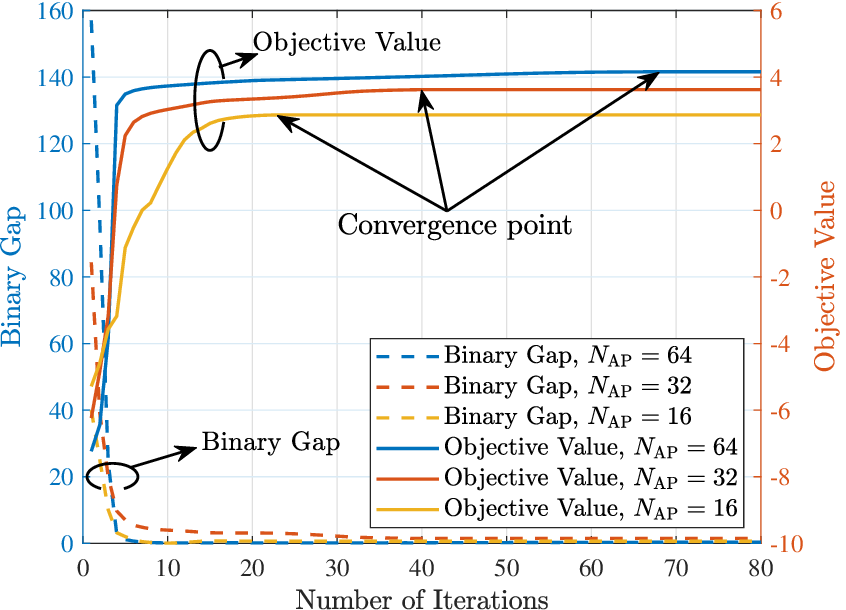}
    \vspace{-0.2cm}
    \caption{The convergence curve of \textbf{Algorithm 1} ($N_{\mathrm{AP}}M_{\mathrm t}=512, K_{\mathrm u}=10, T_{\mathrm{g}}=2, \lambda=10$).}
    \label{fig:fig_0}
    \vspace{-0.2cm}
\end{figure}

\subsection{Convergence Performance}
Fig.~\ref{fig:fig_0} illustrates the convergence behavior of \textbf{Algorithm~\ref{alg1}} for different AP setups. To solve the optimization problem~\eqref{eq:P3}, the initial mode selection vector is set to $a_p^{(0)}=0.5, \forall p\in\mathcal{N}$, with a penalty factor of $\lambda=10$. To accommodate negative values arising from the penalty term, the objective value is presented in a logarithmic form as $\text{sgn}(1+\text{obj})\log_2(|1+\text{obj}|)$, where $\text{sgn}(\cdot)$ represents the sign function. Here, $\text{obj}\triangleq z-g^{(i)}$ and the binary gap is defined as $g^{(i)}\triangleq \lambda \sum\nolimits_{p\in\mathcal{N}}a_p-a_p^{(i)}\big(2a_p-a_p^{(i)}\big)$. The results demonstrate that the binary gap rapidly decreases to 0, which indicates the mode selection vector converges to a binary state, i.e., $a_p \approx a_p^{(i)} \in\{0,1\}$. This confirms the effectiveness of the proposed AP mode selection scheme. At the convergence point, we have the objective value $\text{obj} \approx \log_2(1+z)$, where the auxiliary variable is $z=\text{min}_{q\in\mathcal{Q}}\,\text{SINR}^{(\mathrm c)}_q$, implying the value of the objective function converges to the minimum per-user SE.

\begin{figure}[!t]
    \centering
    \includegraphics[width=2.9 in]{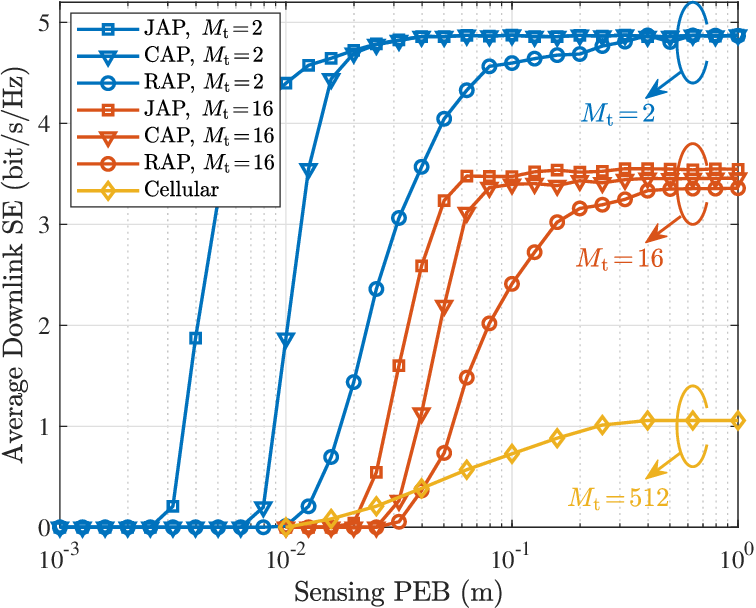}
    \vspace{-0.2cm}
    \caption{Tradeoff between the SE and the sensing PEB constraint in both cellular and CF systems ($N_{\mathrm{AP}}M_{\mathrm t}=512, K_{\mathrm u}=10, T_{\mathrm{g}}=2$).}
    \label{fig:fig_5}
    \vspace{-0.2cm}
\end{figure}

\begin{table}[!t]
\centering
\caption{\footnotesize Execution Time Comparison Between the AP Selection Schemes}
\label{tab3}
\renewcommand\arraystretch{1.5}{
    \centering
    \begin{tabular}{|c|c|c|c|c|}
    \hline
        \multirow{2}{*}{\diagbox[innerleftsep=2.4pt, innerrightsep=2pt, width=2.2cm, height=0.8cm]{\raisebox{-4pt}{\textbf{Scheme}}}{\raisebox{2pt}{\textbf{Time \![s]}}}{\raisebox{2.5pt}{\textbf{Setup}}}} & \multicolumn{4}{c|}{\textbf{Number of APs}} \\ \cline{2-5} & $\!N_{\hspace {-0.05pc}\mathrm{AP}}\!=\!16\!$ & $\!N_{\hspace {-0.05pc}\mathrm{AP}}\!=\!32\!$ & $\!N_{\hspace {-0.05pc}\mathrm{AP}}\!=\!64\!$ & $\!N_{\hspace {-0.05pc}\mathrm{AP}}\!=\!128\!$ \\ \hline
        \textbf{JAP} & 22.02 & 79.12 & 309.37 & 1129.86 \\ \hline
        \textbf{CAP} & 6.06 & 11.34 & 21.94 & 28.17  \\ \hline
    \end{tabular}}
    \vspace{-0.3cm}
\end{table}

\subsection{Tradeoff between Communication and Sensing}
The tradeoff between the communication SE and the sensing PEB constraint under different antenna configurations and AP selection schemes is illustrated in Fig.~\ref{fig:fig_5}. For a fair comparison, the total number of antennas $N_{\mathrm{AP}}M_{\mathrm t}=512$ remains fixed. The results indicate that relaxing the sensing PEB constraints leads to an increase in average communication SE. Compared to conventional cellular ISAC with $M_{\mathrm t}=\text{512}$, the CF-ISAC system significantly enhances SE performance, as denser AP deployments shorten the distances to both users and targets, thereby mitigating signal fading for both communication and sensing. Moreover, the proposed \emph{JAP} scheme provides better ISAC performance with a higher cost of algorithm complexity, while the \emph{CAP} scheme achieves a considerable tradeoff between complexity and  performance. This is validated by the execution time comparison as presented in Table~\ref{tab3}, conducted on an Intel\textsuperscript{\textregistered} Core\textsuperscript{\texttrademark} i9-14900KF CPU. It can be observed that the execution time of the \emph{JAP} scheme grows more rapidly than that of the \emph{CAP} scheme as $N_{\mathrm{AP}}$ increases.


\begin{figure}[!t]
    \centering
    \includegraphics[width=3 in]{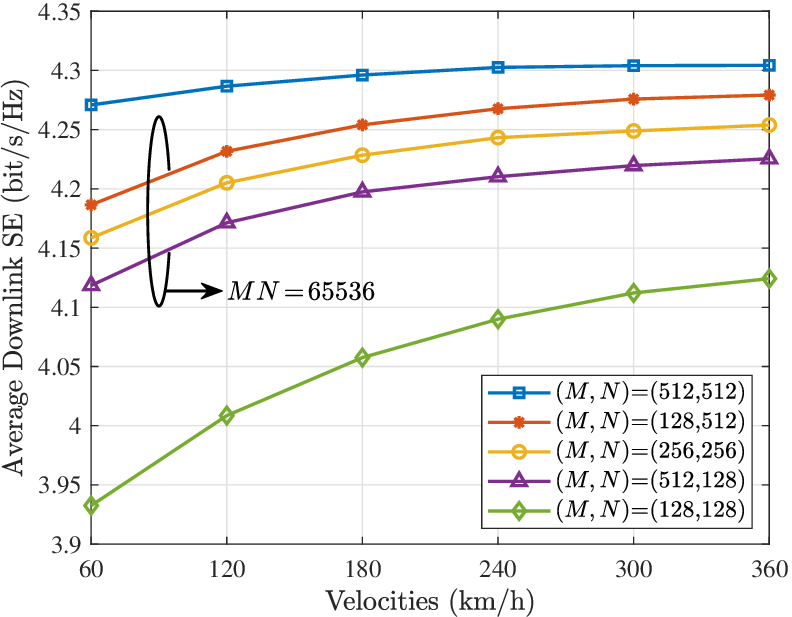}
    \vspace{-0.2cm}
    \caption{The average per-user communication SE versus different user and target velocities ($N_{\mathrm{AP}}=32, K_{\mathrm u}=10, T_{\mathrm{g}}=2$).}
    \label{fig:fig_6}
    \vspace{-0.2cm}
\end{figure}

\subsection{Impact of Mobility and OTFS parameters}
Fig.~\ref{fig:fig_6} investigates the system performance under varying user and target velocities for different numbers of DD grids $(M,N)$. It can be noted that the SE performance gradually improves as velocity increases. The underlying reason is that more distinct paths can be resolved in the Doppler domain as velocity increases, leading to better system performance. The results also demonstrate enhanced SE performance as $M$ and $N$ increase, with the growth in $N$ having a more significant impact. This enhancement is attributed to increased Fisher information provided by finer DD grids, and the $N$-associated Doppler resolution plays a more critical role in positioning under the simulation settings. Nevertheless, increasing the number of DD grids also amplifies computational complexity, revealing the complexity-performance tradeoff from a new perspective.

\begin{figure}[!t]
    \centering
    \includegraphics[width=2.9 in]{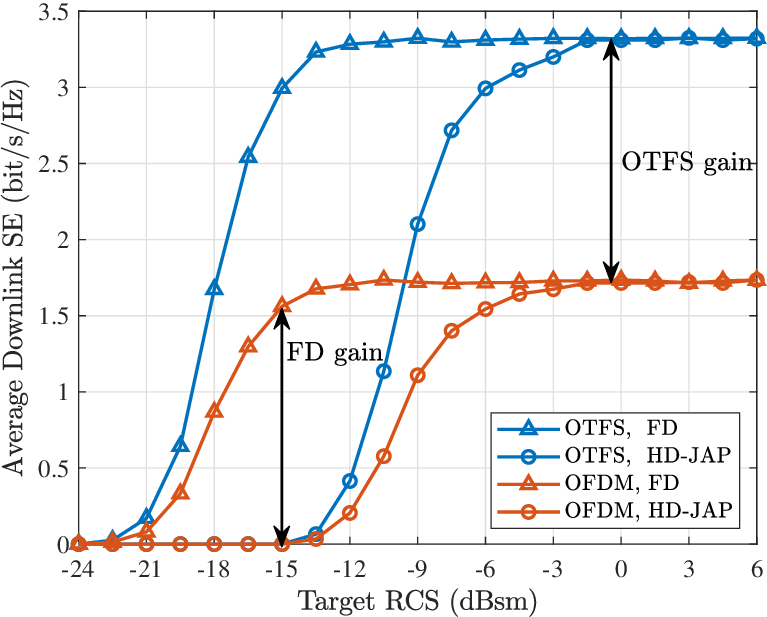}
    \vspace{-0.2cm}
    \caption{The average per-user SE versus RCS variance for different ISAC signals and AP configurations. The JAP algorithm is used to select AP mode in HD configuration ($N_{\mathrm{AP}}=32, K_{\mathrm u}=10, T_{\mathrm{g}}=2$).}
    \label{fig:fig_7}
    \vspace{-0.2cm}
\end{figure}

\begin{table}[!t]
\centering
\caption{\footnotesize Comparison Between Gap Factors in OTFS and OFDM Signals}
\label{tab4}
\renewcommand\arraystretch{1.5}{
    \centering
    \begin{tabular}{|c|c|c|c|}
    \hline
        \textbf{ISAC signal} & \textbf{CP overhead} & \textbf{Pre-log factor} & \makecell{\textbf{Downlink SE} \\ {at RCS $=0$ dBsm}} \\ \hline
        OFDM & 50.0\%  & 0.500 & 1.71 (bit/s/Hz) \\ \hline
        OTFS & 0.78\%  & 0.992 & 3.31 (bit/s/Hz) \\ \hline
    \end{tabular}}
\vspace{-0.3cm}
\end{table}

\subsection{Impact of ISAC Signal and Full-Duplex Configuration}
Finally, the performance gap between the systems based on the OTFS and OFDM signals at different target RCS values, for both full-duplex (FD) and half-duplex (HD) AP configurations, is shown in Fig.~\ref{fig:fig_7}. The results show that the OTFS signal achieves a nearly twofold higher SE than its OFDM counterpart. The underlying reason is that the OFDM signal requires a CP added before each symbol within a data block to mitigate inter-symbol interference (ISI), resulting in a total number of $N$ CPs, whereas the OTFS signal uses only a single CP for the entire data block. As detailed in Table~\ref{tab4}, the OFDM signal's large CP overhead significantly reduces the pre-log factor $\omega$ and the communication SE, thereby underscoring considerable potential of the OTFS signal for broadband systems. By enabling additional links for both communication and multi-static sensing, the FD configuration achieves notable SE improvements. However, effective cancellation of the severe self-interference inherent in the FD configuration remains challenging~\cite{Ren2025Two}.

\section{Conclusion}
This paper studied the multi-static position estimation in the CF-ISAC systems employing the OTFS signal, concurrently analyzing and optimizing the system performance. Specifically, by directly estimating the target positions in a common reference system, an ML position estimation scheme with a smaller search space was proposed. A closed-form CRLB expression and its low-complexity approximation for target position estimation were derived, and a universal structure of multi-static sensing CRLB was summarized to support other ISAC signals. Moreover, a joint optimization algorithm and a low-complexity decomposition method were proposed to solve the joint AP mode selection and power allocation design problem.

The numerical results validated the derived PEB expression and approximation, which clearly showed the coordination gain of ISAC signals and paved the way for system optimization. The proposed ML position estimation scheme achieved promising performance that closely approached the theoretical PEB at high sensing SNRs, demonstrating its effectiveness. A remarkable performance enhancement of the joint optimization algorithm over a random AP mode benchmark was presented, and a considerable tradeoff between algorithm complexity and ISAC performance was achieved by the decomposition method. In addition, the superiority of the OTFS signal over the OFDM signal was analyzed due to different CP mechanisms, and the vital role of the AP antenna array directions in elevating the sensing performance was demonstrated. This finding indicated an interesting focus for future ISAC research, namely, employing digital receiving beamforming instead of mechanical receiving AP array steering.

{\appendices
\section{Entries of the FIM in \texorpdfstring{\eqref{eq:FIM}}{\ref{eq:FIM}}}
\label{app:FIM_Entries}
\begingroup\makeatletter\def\f@size{9}\check@mathfonts
\allowdisplaybreaks
\vspace{-0.4cm}
\begin{align*}
F_{\omega_{p,r,v}^{\mathrm{t}},\omega_{p,r,v}^{\mathrm{t}}}&=\frac{2}{\sigma_w^2}\Re\{(\beta_{p}^{*}\beta_{p}^{\phantom{*}}\mathbf{h}_{vr}^{\dagger}\mathbf{h}_{vr}^{\phantom{\dagger}})(\dot{\mathbf{h}}_{pv}^{\dagger}\mathbf{V}_{p}^{\phantom{\dagger}}\dot{\mathbf{h}}_{pv}^{\phantom{\dagger}})R_{p,r,v}^{(0,0)}\}\\
F_{\tau_{p,r,v},\tau_{p,r,v}}&=\frac{2}{\sigma_w^2}\Re\{(\beta_{p}^{*}\beta_{p}^{\phantom{*}}\mathbf{h}_{vr}^{\dagger}\mathbf{h}_{vr}^{\phantom{\dagger}})(\mathbf{h}_{pv}^{\dagger}\mathbf{V}_{p}^{\phantom{\dagger}}\mathbf{h}_{pv}^{\phantom{\dagger}})R_{p,r,v}^{(2,0)}\}\\
F_{\nu_{p,r,v},\nu_{p,r,v}}&=\frac{2}{\sigma_w^2}\Re\{(\beta_{p}^{*}\beta_{p}^{\phantom{*}}\mathbf{h}_{vr}^{\dagger}\mathbf{h}_{vr}^{\phantom{\dagger}})(\mathbf{h}_{pv}^{\dagger}\mathbf{V}_{p}^{\phantom{\dagger}}\mathbf{h}_{pv}^{\phantom{\dagger}})R_{p,r,v}^{(0,2)}\}\\
F_{\beta_{p,r,v}^{(\mathrm{R})},\beta_{p,r,v}^{(\mathrm{R})}}&=\frac{2}{\sigma_w^2}\Re\{(\mathbf{h}_{vr}^{\dagger}\mathbf{h}_{vr}^{\phantom{\dagger}})(\mathbf{h}_{pv}^{\dagger}\mathbf{V}_{p}^{\phantom{\dagger}}\mathbf{h}_{pv}^{\phantom{\dagger}})R_{p,r,v}^{(0,0)}\}\\
F_{\beta_{p,r,v}^{(\mathrm{I})},\beta_{p,r,v}^{(\mathrm{I})}}&=\frac{2}{\sigma_w^2}\Re\{(\mathbf{h}_{vr}^{\dagger}\mathbf{h}_{vr}^{\phantom{\dagger}})(\mathbf{h}_{pv}^{\dagger}\mathbf{V}_{p}^{\phantom{\dagger}}\mathbf{h}_{pv}^{\phantom{\dagger}})R_{p,r,v}^{(0,0)}\}\\
F_{\omega_{p,r,v}^{\mathrm{r}},\omega_{p,r,v}^{\mathrm{t}}}&=-\frac{2}{\sigma_w^2}\Re\{(\beta_{p}^{*}\beta_{p}^{\phantom{*}}\dot{\mathbf{h}}_{vr}^{\dagger}\mathbf{h}_{vr}^{\phantom{\dagger}})(\dot{\mathbf{h}}_{pv}^{\dagger}\mathbf{V}_{p}^{\phantom{\dagger}}\mathbf{h}_{pv}^{\phantom{\dagger}})R_{p,r,v}^{(0,0)}\}\\
F_{\omega_{p,r,v}^{\mathrm{r}},\tau_{p,r,v}}&=\frac{2}{\sigma_w^2}\Re\{j(\beta_{p}^{*}\beta_{p}^{\phantom{*}}\dot{\mathbf{h}}_{vr}^{\dagger}\mathbf{h}_{vr}^{\phantom{\dagger}})(\mathbf{h}_{pv}^{\dagger}\mathbf{V}_{p}^{\phantom{\dagger}}\mathbf{h}_{pv}^{\phantom{\dagger}})R_{p,r,v}^{(1,0)}\}\\
F_{\omega_{p,r,v}^{\mathrm{r}},\nu_{p,r,v}}&=\frac{2}{\sigma_w^2}\Re\{j(\beta_{p}^{*}\beta_{p}^{\phantom{*}}\dot{\mathbf{h}}_{vr}^{\dagger}\mathbf{h}_{vr}^{\phantom{\dagger}})(\mathbf{h}_{pv}^{\dagger}\mathbf{V}_{p}^{\phantom{\dagger}}\mathbf{h}_{pv}^{\phantom{\dagger}})R_{p,r,v}^{(0,1)}\}\\
F_{\omega_{p,r,v}^{\mathrm{r}},\beta_{p,r,v}^{(\mathrm{R})}}&=\frac{2}{\sigma_w^2}\Re\{j(\beta_{p}^{*}\dot{\mathbf{h}}_{vr}^{\dagger}\mathbf{h}_{vr}^{\phantom{\dagger}})(\mathbf{h}_{pv}^{\dagger}\mathbf{V}_{p}^{\phantom{\dagger}}\mathbf{h}_{pv}^{\phantom{\dagger}})R_{p,r,v}^{(0,0)}\}\\
F_{\omega_{p,r,v}^{\mathrm{r}},\beta_{p,r,v}^{(\mathrm{I})}}&=-\frac{2}{\sigma_w^2}\Re\{(\beta_{p}^{*}\dot{\mathbf{h}}_{vr}^{\dagger}\mathbf{h}_{vr}^{\phantom{\dagger}})(\mathbf{h}_{pv}^{\dagger}\mathbf{V}_{p}^{\phantom{\dagger}}\mathbf{h}_{pv}^{\phantom{\dagger}})R_{p,r,v}^{(0,0)}\}\\
F_{\omega_{p,r,v}^{\mathrm{t}},\tau_{p,r,v}}&=-\frac{2}{\sigma_w^2}\Re\{j(\beta_{p}^{*}\beta_{p}^{\phantom{*}}\mathbf{h}_{vr}^{\dagger}\mathbf{h}_{vr}^{\phantom{\dagger}})(\mathbf{h}_{pv}^{\dagger}\mathbf{V}_{p}^{\phantom{\dagger}}\dot{\mathbf{h}}_{pv}^{\phantom{\dagger}})R_{p,r,v}^{(1,0)}\}\\
F_{\omega_{p,r,v}^{\mathrm{t}},\nu_{p,r,v}}&=-\frac{2}{\sigma_w^2}\Re\{j(\beta_{p}^{*}\beta_{p}^{\phantom{*}}\mathbf{h}_{vr}^{\dagger}\mathbf{h}_{vr}^{\phantom{\dagger}})(\mathbf{h}_{pv}^{\dagger}\mathbf{V}_{p}^{\phantom{\dagger}}\dot{\mathbf{h}}_{pv}^{\phantom{\dagger}})R_{p,r,v}^{(0,1)}\}\\
F_{\omega_{p,r,v}^{\mathrm{t}},\beta_{p,r,v}^{(\mathrm{R})}}&=-\frac{2}{\sigma_w^2}\Re\{j(\beta_{p}^{*}\mathbf{h}_{vr}^{\dagger}\mathbf{h}_{vr}^{\phantom{\dagger}})(\mathbf{h}_{pv}^{\dagger}\mathbf{V}_{p}^{\phantom{\dagger}}\dot{\mathbf{h}}_{pv}^{\phantom{\dagger}})R_{p,r,v}^{(0,0)}\}\\
F_{\omega_{p,r,v}^{\mathrm{t}},\beta_{p,r,v}^{(\mathrm{I})}}&=\frac{2}{\sigma_w^2}\Re\{(\beta_{p}^{*}\mathbf{h}_{vr}^{\dagger}\mathbf{h}_{vr}^{\phantom{\dagger}})(\mathbf{h}_{pv}^{\dagger}\mathbf{V}_{p}^{\phantom{\dagger}}\dot{\mathbf{h}}_{pv}^{\phantom{\dagger}})R_{p,r,v}^{(0,0)}\}\\
F_{\tau_{p,r,v},\nu_{p,r,v}}&=\frac{2}{\sigma_w^2}\Re\{(\beta_{p}^{*}\beta_{p}^{\phantom{*}}\mathbf{h}_{vr}^{\dagger}\mathbf{h}_{vr}^{\phantom{\dagger}})(\mathbf{h}_{pv}^{\dagger}\mathbf{V}_{p}^{\phantom{\dagger}}\mathbf{h}_{pv}^{\phantom{\dagger}})R_{p,r,v}^{(1,1)}\}\\
F_{\tau_{p,r,v},\beta_{p,r,v}^{(\mathrm{R})}}&=\frac{2}{\sigma_w^2}\Re\{(\beta_{p}^{*}\mathbf{h}_{vr}^{\dagger}\mathbf{h}_{vr}^{\phantom{\dagger}})(\mathbf{h}_{pv}^{\dagger}\mathbf{V}_{p}^{\phantom{\dagger}}\mathbf{h}_{pv}^{\phantom{\dagger}})R_{p,r,v}^{*(1,0)}\}\\
F_{\tau_{p,r,v},\beta_{p,r,v}^{(\mathrm{I})}}&=\frac{2}{\sigma_w^2}\Re\{j(\beta_{p}^{*}\mathbf{h}_{vr}^{\dagger}\mathbf{h}_{vr}^{\phantom{\dagger}})(\mathbf{h}_{pv}^{\dagger}\mathbf{V}_{p}^{\phantom{\dagger}}\mathbf{h}_{pv}^{\phantom{\dagger}})R_{p,r,v}^{*(1,0)}\}\\
F_{\nu_{p,r,v},\beta_{p,r,v}^{(\mathrm{R})}}&=\frac{2}{\sigma_w^2}\Re\{(\beta_{p}^{*}\mathbf{h}_{vr}^{\dagger}\mathbf{h}_{vr}^{\phantom{\dagger}})(\mathbf{h}_{pv}^{\dagger}\mathbf{V}_{p}^{\phantom{\dagger}}\mathbf{h}_{pv}^{\phantom{\dagger}})R_{p,r,v}^{*(0,1)}\}\\
F_{\nu_{p,r,v},\beta_{p,r,v}^{(\mathrm{I})}}&=\frac{2}{\sigma_w^2}\Re\{j(\beta_{p}^{*}\mathbf{h}_{vr}^{\dagger}\mathbf{h}_{vr}^{\phantom{\dagger}})(\mathbf{h}_{pv}^{\dagger}\mathbf{V}_{p}^{\phantom{\dagger}}\mathbf{h}_{pv}^{\phantom{\dagger}})R_{p,r,v}^{*(0,1)}\}\\
F_{\beta_{p,r,v}^{(\mathrm{R})},\beta_{p,r,v}^{(\mathrm{I})}}&=\frac{2}{\sigma_w^2}\Re\{j(\mathbf{h}_{vr}^{\dagger}\mathbf{h}_{vr}^{\phantom{\dagger}})(\mathbf{h}_{pv}^{\dagger}\mathbf{V}_{p}^{\phantom{\dagger}}\mathbf{h}_{pv}^{\phantom{\dagger}})R_{p,r,v}^{(0,0)}\}
\end{align*}
\endgroup
\vspace{-1mm}
\vspace{-1mm}
\vspace{-1mm}
\vspace{-1mm}
\vspace{-1mm}
\vspace{-1mm}
\vspace{-1mm}

\section{Proof of Theorem \ref{thm1}}
\label{app:thm1}

Before deriving the approximation FIM {\small$\hat{\mathbf{F}}_{\mathbf{p}_{p,r,v}}$}, we define the short notations used in~\eqref{eq:FIM_example} as follows
\vspace{-0.5mm}
\begingroup\makeatletter\def\f@size{9}\check@mathfonts
\allowdisplaybreaks
\begin{subequations}\label{eq:R_original}
\begin{align}
\hspace {-0.25pc}R_{p,r,v}^{(0,0)}\!=&\!\sum_{k=0}^{N-1}\sum_{l=0}^{M-1}\sum_{k^{\prime}=0}^{N-1} \sum_{l^{\prime}=0}^{M-1}  \left(\Psi_{k,k^{\prime},l,l^{\prime}}^{p,r,v}\right)^{*}\Psi_{k,k^{\prime},l,l^{\prime}}^{p,r,v},\!\label{eq:R00}\\
\hspace {-0.25pc}R_{p,r,v}^{(1,0)}\!=&\!\sum_{k=0}^{N-1}\sum_{l=0}^{M-1}\sum_{k^{\prime}=0}^{N-1} \sum_{l^{\prime}=0}^{M-1}  \left(\Psi_{k,k^{\prime},l,l^{\prime}}^{p,r,v}\right)^{*}\frac{\partial \Psi_{k,k^{\prime},l,l^{\prime}}^{p,r,v}}{\partial \tau_{p,r,v}},\!\label{eq:R10}\\
\hspace {-0.25pc}R_{p,r,v}^{(0,1)}\!=&\!\sum_{k=0}^{N-1}\sum_{l=0}^{M-1}\sum_{k^{\prime}=0}^{N-1} \sum_{l^{\prime}=0}^{M-1}  \left(\Psi_{k,k^{\prime},l,l^{\prime}}^{p,r,v}\right)^{*}\frac{\partial \Psi_{k,k^{\prime},l,l^{\prime}}^{p,r,v}}{\partial \nu_{p,r,v}},\!\\
\hspace {-0.25pc}R_{p,r,v}^{(1,1)}\!=&\!\sum_{k=0}^{N-1}\sum_{l=0}^{M-1}\sum_{k^{\prime}=0}^{N-1} \sum_{l^{\prime}=0}^{M-1}  \left(\frac{\partial \Psi_{k,k^{\prime},l,l^{\prime}}^{p,r,v}}{\partial \tau_{p,r,v}}\right)^{*}\frac{\partial \Psi_{k,k^{\prime},l,l^{\prime}}^{p,r,v}}{\partial \nu_{p,r,v}},\!\\
\hspace {-0.25pc}R_{p,r,v}^{(2,0)}\!=&\!\sum_{k=0}^{N-1}\sum_{l=0}^{M-1}\sum_{k^{\prime}=0}^{N-1} \sum_{l^{\prime}=0}^{M-1}  \left(\frac{\partial \Psi_{k,k^{\prime},l,l^{\prime}}^{p,r,v}}{\partial \tau_{p,r,v}}\right)^{*}\frac{\partial \Psi_{k,k^{\prime},l,l^{\prime}}^{p,r,v}}{\partial \tau_{p,r,v}},\!\\
\hspace {-0.25pc}R_{p,r,v}^{(0,2)}\!=&\!\sum_{k=0}^{N-1}\sum_{l=0}^{M-1}\sum_{k^{\prime}=0}^{N-1} \sum_{l^{\prime}=0}^{M-1}  \left(\frac{\partial \Psi_{k,k^{\prime},l,l^{\prime}}^{p,r,v}}{\partial \nu_{p,r,v}}\right)^{*}\frac{\partial \Psi_{k,k^{\prime},l,l^{\prime}}^{p,r,v}}{\partial \nu_{p,r,v}},\!\label{eq:R02}
\end{align}
\end{subequations}
\endgroup
\vspace{-1mm}
\vspace{-1.5mm}

\noindent where the partial derivatives of $\boldsymbol{\Psi}$ with respect to the channel parameters $\tau$ and $\nu$, ignoring indices $p$, $r$ and $v$, are given by~\cite{Dehkordi2023Beam}
\vspace{-1mm}
\vspace{-1mm}
\begingroup\makeatletter\def\f@size{9}\check@mathfonts
\begin{equation}
\begin{aligned}
\frac{\partial \Psi_{k,k^{\prime},l,l^{\prime}}}{\partial \tau}=&\frac{j 2 \pi \Delta f}{N M} \mathbf{1}_N^\mathrm{T} \boldsymbol{\alpha}_{k, k^{\prime}}(\nu) \mathbf{c}_M^\mathrm{T} \boldsymbol{\beta}_{k^{\prime},l,l^{\prime}}(\nu, \tau),\\
\frac{\partial \Psi_{k,k^{\prime},l,l^{\prime}}}{\partial \nu}=&\frac{j 2 \pi}{N M}\big[T \mathbf{c}_N^\mathrm{T} \boldsymbol{\alpha}_{k,k^{\prime}}(\nu) \mathbf{1}_M^\mathrm{T} \boldsymbol{\beta}_{k^{\prime},l,l^{\prime}}(\nu, \tau) \\& 
+ g(l) \mathbf{1}_N^\mathrm{T} \boldsymbol{\alpha}_{k,k^{\prime}}(\nu) \mathbf{1}_M^\mathrm{T} \boldsymbol{\beta}_{k^{\prime},l,l^{\prime}}(\nu, \tau)\big],
\end{aligned}
\end{equation}
\endgroup
\vspace{-1mm}
\vspace{-1mm}

\begingroup\makeatletter\def\f@size{9}\check@mathfonts
\noindent where $g(l)\!=\!\frac{l}{M\Delta f}$ for $l\!\in\!\mathcal{L}_{\mathrm{ICI}}(\tau)$ and $g(l)\!=\!\frac{l}{M\Delta f}-T$ for $l\!\in\!\mathcal{L}_{\mathrm{ISI}}(\tau)$; meanwhile, the vectors $\boldsymbol{\alpha}_{k,k^{\prime}}\!$ and $\boldsymbol{\beta}_{k^{\prime},l,l^{\prime}}\!$ are defined as $\boldsymbol{\alpha}_{k,k^{\prime}}(\nu)\!=\![\alpha_{0,k,k^{\prime}}(\nu),\ldots,\alpha_{N-1,k,k^{\prime}}(\nu)]^\mathrm{T}\!\in\!\mathbb{C}^{N\times1}$ and $\boldsymbol{\beta}_{k^{\prime},l,l^{\prime}}(\nu,\tau)\!=\![\beta_{0,k^{\prime},l,l^{\prime}}(\nu,\tau),\ldots,\beta_{M-1,k^{\prime},l,l^{\prime}}(\nu,\tau)]^\mathrm{T}\!\in\!\mathbb{C}^{M\times1}$, respectively.
\endgroup

\begingroup\makeatletter\def\f@size{9}\check@mathfonts
Next, we proceed with the derivation of $R_{p,r,v}^{(0,0)}$. By substituting~\eqref{eq:Psi} into~\eqref{eq:R00}, the $R_{p,r,v}^{(0,0)}$ can be derived as
\vspace{-1mm}
\endgroup
\begin{subequations}\label{eq:R_derived}
\begingroup\makeatletter\def\f@size{9}\check@mathfonts
\begin{align}
\hspace {-0.5pc}R_{p,r,v}^{(0,0)}=&\frac{1}{M^2}\!\sum_{l=0}^{M-1}\sum_{l^{\prime}=0}^{M-1}\sum_{m^{\prime}=0}^{M-1}\sum_{m^{\prime\prime}=0}^{M-1}\!e^{-j2\pi\frac{l^{\prime}-l+\tau M\Delta f}{M}(m^{\prime}\!-m^{\prime\prime})}\!\!\nonumber\\
&\times\frac{1}{N^2}\!\sum_{k=0}^{N-1}\sum_{k^{\prime}=0}^{N-1}\sum_{n^{\prime}=0}^{N-1}\sum_{n^{\prime\prime}=0}^{N-1}\!e^{-j2\pi\frac{k^{\prime}-k+\nu NT}{N}(n^{\prime}\!-n^{\prime\prime})}\!\!\nonumber\\
\stackrel {(a)}{=}&\frac{1}{M^2N^2}\times M^3N^3=MN,
\end{align}
\endgroup
\vspace{-1mm}
\vspace{-1mm}
\vspace{-1mm}
\vspace{-1mm}
\vspace{-1mm}

\noindent where in (a), we note that the sum is nonzero only when $m^{\prime}\!-m^{\prime\prime}=0$ and $n^{\prime}\!-n^{\prime\prime}=0$. The remaining terms~\eqref{eq:R10}-\eqref{eq:R02} can be similarly obtained as follows
\vspace{-1mm}
\begingroup\makeatletter\def\f@size{9}\check@mathfonts
\begin{align}
\hspace {-0.5pc}R_{p,r,v}^{(1,0)}\!=&j\pi\Delta f\left(M\!-\!1\right)MN\\
\hspace {-0.5pc}R_{p,r,v}^{(0,1)}\!=&j\pi\left[T\left(N\!-\!1\right)MN+2N\!\sum\nolimits_{l=0}^{M-1}\!\!g(l)\right]\\       
\hspace {-0.5pc}R_{p,r,v}^{(1,1)}\!=&\pi^2\left(M\!-\!1\right)N\!\left[\left(N\!-\!1\right)M+2\Delta f\!\sum\nolimits_{l=0}^{M-1}\!\!g(l)\right]\!\!\\
\hspace {-0.5pc}R_{p,r,v}^{(2,0)}\!=&\frac{\left(2\pi\Delta f\right)^2\!\left(M\!-\!1\right)MN\left(2M\!-\!1\right)}{6}\\
\hspace {-0.5pc}R_{p,r,v}^{(0,2)}\!=&\frac{\left(2\pi T\right)^2\!\left(N\!-\!1\right)MN\left(2N\!-\!1\right)}{6}+\left(2\pi\right)^2\!N\!\sum\nolimits_{l=0}^{M\!-\!1}\!\!\!g^2(l)\nonumber\\&+\left(2\pi\right)^2T\left(N\!-\!1\right)N\!\sum\nolimits_{l=0}^{M\!-\!1}\!\!g(l)
\end{align}
\endgroup
\end{subequations}
\vspace{-1mm}
\vspace{-1mm}
\vspace{-1mm}
\vspace{-1mm}

\begingroup\makeatletter\def\f@size{9.5}\check@mathfonts
\noindent Further, using the identity $\left(\mathbf{a}\odot \mathbf{b}\right)^{\dagger}\!\left(\mathbf{c}\odot \mathbf{d}\right)\!=\!\left(\mathbf{a}\odot \mathbf{d}\right)^{\dagger}\!\left(\mathbf{c}\odot \mathbf{b}\right)$, and substituting {\small$\mathbf{V}_{p}\approx\eta_{pv}\hat{\mathbf{h}}_{pv}^{\phantom{\dagger}}\hat{\mathbf{h}}_{pv}^{\dagger}$} yield
\endgroup
\vspace{-1mm}
\vspace{-1mm}
\begingroup\makeatletter\def\f@size{9}\check@mathfonts
\begin{equation}
\begin{aligned}
\hspace {-0.4pc}&\mathbf{h}_{vr}^{\dagger}\mathbf{h}_{vr}^{\phantom{\dagger}}\!\hspace{-0.1mm}=M_{\mathrm t},\qquad\qquad\qquad\quad\ \ \!\hspace{-0.1mm}
\mathbf{h}_{pv}^{\dagger}\!\mathbf{V}_{\!p}^{\phantom{\dagger}}\hspace{-0.2mm}\mathbf{h}_{pv}^{\phantom{\dagger}}\!\approx\hspace{-0.2mm}\eta_{pv}M_{\mathrm t},
\\
\hspace {-0.4pc}&\dot{\mathbf{h}}_{vr}^{\dagger}\mathbf{h}_{vr}^{\phantom{\dagger}}\!\hspace{-0.1mm}=\mathbf{h}_{vr}^{\dagger}\dot{\mathbf{h}}_{vr}^{\phantom{\dagger}}\!=\!\frac{M_{\mathrm t}\!\left(M_{\mathrm t}\!-\!1\right)}{2},\quad\!\!\!\!\hspace{-0.2mm}\dot{\mathbf{h}}_{pv}^{\dagger}\!\mathbf{V}_{\!p}^{\phantom{\dagger}}\hspace{-0.2mm}\mathbf{h}_{pv}^{\phantom{\dagger}}\!\approx\hspace{-0.2mm}\eta_{pv}\frac{M_{\mathrm t}\!\left(M_{\mathrm t}\!-\!1\right)}{2}\!,\!\!\!
\\
\hspace {-0.4pc}&\dot{\mathbf{h}}_{vr}^{\dagger}\dot{\mathbf{h}}_{vr}^{\phantom{\dagger}}\!\hspace{-0.1mm}=\!\frac{M_{\mathrm t}\!\left(M_{\mathrm t}\!-\!1\right)\!\left(2M_{\mathrm t}\!-\!1\right)}{6},\quad\!\!\!
\dot{\mathbf{h}}_{pv}^{\dagger}\!\mathbf{V}_{\!p}^{\phantom{\dagger}}\hspace{-0.2mm}\dot{\mathbf{h}}_{pv}^{\phantom{\dagger}}\!\approx\hspace{-0.2mm}\eta_{pv}\frac{M_{\mathrm t}\!\left(M_{\mathrm t}\!-\!1\right)^2}{4}\!.\!\!\!\!\!\!\!\!\!
\label{eq:h_derived}
\end{aligned}
\end{equation}
\endgroup
\vspace{-1mm}
\vspace{-1mm}

\noindent Then, by substituting~\eqref{eq:R_derived} and~\eqref{eq:h_derived} into~\eqref{eq:FIM_example}, the equivalent FIM matrix in~\eqref{eq:EFIM} is obtained as
\vspace{-1mm}
\vspace{-0.5mm}
\begingroup\makeatletter\def\f@size{9.5}\check@mathfonts
\begin{equation}
\mathbf{F}_{\boldsymbol{\theta}_{p,r,v}^{(1)}}^{\mathrm{e}}=\frac{2|\beta_{p}|^{2}\eta_{pv}}{\sigma_w^2}\operatorname{diag}\{d_{11},d_{22},d_{33},d_{44}\},
\label{eq:EFIM_derived}
\end{equation}
\vspace{-1mm}
\vspace{-1mm}
\vspace{-1mm}
\vspace{-1mm}
\vspace{-1mm}

\noindent where
\vspace{-1mm}
\vspace{-1mm}
\begin{align*}
\hspace {-0.5pc}d_{11}=&\frac{M_{\mathrm t}\!\left(M_{\mathrm t}\!-\!1\right)\!\left(2M_{\mathrm t}\!-\!1\right)MN}{6} - \frac{M_{\mathrm t}\!\left(M_{\mathrm t}\!-\!1\right)^2\!MN}{4}, d_{22}=0, \\
\hspace {-0.5pc}d_{33}=&R_{p,r,v}^{(2,0)}+\big(R_{p,r,v}^{(1,0)}\big)^2/MN,\ d_{44}=R_{p,r,v}^{(0,2)}+\big(R_{p,r,v}^{(0,1)}\big)^2/MN.\!\!\!
\label{eq:d_derived}
\end{align*}
\endgroup
\vspace{-1mm}
\vspace{-1mm}
\vspace{-1mm}
\vspace{-1mm}
\vspace{-1mm}

Finally, by substituting~\eqref{eq:EFIM_derived} into~\eqref{eq:positionFIM}, the desired result in~\eqref{eq:approxPositionFIM} is obtained following a series of algebraic manipulations.

\section{Proof of Corollary \ref{cor1}}
\label{app:cor1}
\begingroup\makeatletter\def\f@size{9.5}\check@mathfonts
It can be noted that substituting the OTFS signal with the OFDM signal modifies merely the signal factors within the FIM form~\eqref{eq:FIM_form}, while the RX and TX factors remain unaffected. Therefore, we proceed to derive the revised signal factors $R_{p,r,v}$ under the OFDM signal.

For a fair comparison, we assume that each OFDM symbol duration is $T=T_{\mathrm{cp}}+T_0$, where $T_{\mathrm{cp}}$ and $T_0$ denote the CP and data symbol durations, respectively. Then, the TF domain input-output relationship can be formulated as~\cite{Gaudio2019Performance}
\endgroup
\vspace{-1.5mm}
\begingroup\makeatletter\def\f@size{9.5}\check@mathfonts
\begin{equation}
y[n,m]=\sum\nolimits_{m^{\prime}=0}^{M-1}\Psi_{n,m,m^{\prime}}x[n,m^{\prime}],
\end{equation}
\vspace{-1mm}
\vspace{-1mm}
\vspace{-2.5mm}

\noindent where the effective TF domain channel is given by
\vspace{-1mm}
\begin{equation}\label{eq:Psi_OFDM}
\hspace{-0.4pc}\Psi_{n,m,m^{\prime}}\!=\!\frac{1}{M}e^{j2\pi m^{\prime}\tau\Delta f}e^{j2\pi n\nu T}\!\sum\nolimits_{i=0}^{M-1}\!e^{j2\pi(m^{\prime}-m+\nu T)\frac{i}{M}}\!\!.\!
\end{equation}
\endgroup

\noindent Next, by substituting~\eqref{eq:Psi_OFDM} into~\eqref{eq:R_original} and following similar steps for deriving its OTFS counterpart in~\eqref{eq:R_derived}, the signal factors in the FIM expressions for the OFDM signal can be calculated as
\vspace{-1mm}
\vspace{-1mm}
\vspace{-1mm}
\vspace{-1mm}
\vspace{-1mm}
\vspace{-0.5mm}
\begingroup\makeatletter\def\f@size{9}\check@mathfonts
\begin{subequations}\label{eq:R_OFDM_derived}
\begin{align}
\hspace{-0.5pc}R_{p,r,v}^{(0,0)}\!=&MN\\
R_{p,r,v}^{(1,0)}\!=&j\pi\Delta f\left(M-1\right)MN\\
R_{p,r,v}^{(0,1)}\!=&j\pi\big[T\left(N-1\right)MN+T_0\left(M-1\right)N\big]\\
R_{p,r,v}^{(1,1)}\!=&\pi^2\left(M-1\right)N\big[\left(N-1\right)M+T_0\Delta f\left(M-1\right)\big]\!\\
R_{p,r,v}^{(2,0)}\!=&\frac{\left(2\pi\Delta f\right)^2\left(M\!-\!1\right)M\!N\left(2M\!-\!1\right)}{6}\\
R_{p,r,v}^{(0,2)}\!=&\frac{\left(2\pi T\right)^2\!\left(N\!-\!1\right)\!M\!N\!\left(2N\!-\!1\right)}{6}\!+\!\frac{\left(2\pi T_0\right)^2\!\left(M\!-\!1\right)\!N\!\left(2M\!-\!1\right)}{6M}\nonumber\\&+2\pi^2T_0T\left(N\!-\!1\right)\!N\!\left(M\!-\!1\right)\!.\!\!
\end{align}
\end{subequations}
\endgroup

}

\bibliographystyle{IEEEtran}
\bibliography{IEEEabrv,bibRef}

\end{document}